\documentclass[fleqn,usenatbib]{mnras}
\usepackage[T1]{fontenc}
\usepackage{ae,aecompl}
\usepackage{graphicx}
\usepackage{amsmath}
\usepackage{amssymb}
\usepackage{url}
\graphicspath{{figure/}}

\newcommand{\kms}{\,km\,s$^{-1}$}
\newcommand{\molh}{$\text{H}_2$}
\newcommand{\Msun}{M$_\odot$}
\newcommand{\pc}[2]{pc$^{-2}$}
\newcommand{\Lya}{Ly$\alpha$}
\newcommand{\xion}{$\xi_\text{ion}$}
\newcommand{\Nion}{$\dot{N}_\text{ion}$}

\defcitealias{Yung2019}{Paper I}
\defcitealias{Yung2019a}{Paper II}
\defcitealias{Yung2020a}{Paper IV}

\defcitealias{Somerville2008}{S08} 
\defcitealias{Somerville2012}{S12}
\defcitealias{Popping2014}{PST14}
\defcitealias{Somerville2015}{SPT15}
\defcitealias{Somerville2015a}{SD15}

\defcitealias{Bruzual2003}{BC03}
\defcitealias{Gnedin2011}{GK}
\defcitealias{Bigiel2008}{Big}
\defcitealias{Kuhlen2012}{KF12}
\defcitealias{Wilkins2016a}{W16}
\defcitealias{Finkelstein2019}{F19}

\title[SAM forecasts -- III. LyC production efficiency]{Semi-analytic forecasts for \textit{JWST} -- III. Intrinsic production efficiency of Lyman-continuum radiation}

\author[L. Y. A. Yung et al.]{L. Y. Aaron Yung,$^{1,2}$\thanks{E-mail: yung@physics.rutgers.edu}
Rachel S. Somerville,$^{1,2}$ Gerg\"{o} Popping$^{3}$ 
\newauthor and Steven L. Finkelstein$^{4}$
\\
$^{1}$ Department of Physics and Astronomy, Rutgers University, 136 Frelinghuysen Road, Piscataway, NJ 08854, USA\\
$^{2}$ Center for Computational Astrophysics, Flatiron Institute, 162 5th Ave, New York, NY 10010, USA\\
$^{3}$ European Southern Observatory, Karl-Schwarzschild-Strasse 2, D-85748 Garching, Germany\\
$^{4}$ Department of Astronomy, The University of Texas at Austin, Austin, TX 78712, USA
}

\date{Accepted XXX. Received YYY; in original form ZZZ}

\pubyear{2019}

\begin{document}
\label{firstpage}
\pagerange{\pageref{firstpage}--\pageref{lastpage}}
\maketitle

\begin{abstract}
The \textit{James Webb Space Telescope} (\textit{JWST}) is expected to enable transformational progress in studying galaxy populations in the very early Universe, during the Epoch of Reionization (EoR). A critical parameter for understanding the sources that reionized the Universe is the Lyman-continuum production efficiency, \xion, defined as the rate of production of ionizing photons divided by the intrinsic UV luminosity. In this work, we combine self-consistent star formation and chemical enrichment histories predicted by semi-analytic models of galaxy formation with stellar population synthesis (SPS) models to predict the expected dependence of \xion\ on galaxy properties and cosmic epoch from $z=4$--10. We then explore the sensitivity of the production rate of ionizing photons, \Nion, to the choice of SPS model and the treatment of stellar feedback in our galaxy formation model. We compare our results to those of other simulations, constraints from empirical models, and observations. We find that adopting SPS models that include binary stars predict about a factor of two more ionizing radiation than models that only assume single stellar populations. We find that UV-faint, low-mass galaxies have values of \xion\ about 0.25 dex higher than those of more massive galaxies, but find weak evolution with cosmic time, about 0.2 dex from $z\sim 12$--4 at fixed rest-UV luminosity. We provide predictions of \Nion\ as a function of $M_\text{h}$ and a number of other galaxy properties.
All results presented in this work are available at \url{https://www.simonsfoundation.org/semi-analytic-forecasts-for-jwst/}.
\end{abstract}



\section{Introduction}
The ionization of hydrogen in the intergalactic medium (IGM) is a critical landmark in cosmic history. The onset and duration of this transition is constrained by various kinds of observations, including the polarization of the cosmic microwave background (CMB), the spectra of high-redshift ($z\gtrsim6$) quasars, and the abundance of \Lya\ emitters \citep[e.g.][]{Fan2006, Robertson2013}. Despite the uncertainties in these observational constraints, the astronomical community has reached a broad consensus that the process of hydrogen reionization took place roughly between $z\approx6$--10, which is commonly referred to as the Epoch of Reionization (EoR) \citep{Loeb2001}. However, there has been ongoing tension between the total ionizing photon budget accounted for by all known sources and the time frame of cosmic reionization set by current observations.

Although it is clear that the population of directly observed high-redshift galaxies observed to date alone are insufficient to ionize the IGM \citep[e.g.][]{Bouwens2015a, Finkelstein2015, Robertson2015}, many recent analytic studies have shown that a reasonable extrapolation of the observed galaxy populations during the EoR down to fainter rest-frame UV luminosities (e.g. $M_\text{UV} \sim -13$ may be able to account for most reionization constraints \citep{Finkelstein2012a, Finkelstein2015, Finkelstein2019, Kuhlen2012, Bouwens2015a, Robertson2015, Naidu2019}. This is further supported by cosmological hydrodynamic simulations, many of which are able to satisfy the constraints \citep[e.g.][]{Geil2016, Ocvirk2016, Ocvirk2018, Anderson2017}. However, there are still quite a few remaining uncertainties in the estimation of the production rate of ionizing photons.

The total number of ionizing photons available to reionize the IGM depends on the abundance of star-forming galaxies during the EoR, their intrinsic production efficiency of Lyman-continuum (LyC) radiation, and the fraction of radiation that escapes to the IGM. Each of these three moving parts comes with its own substantial set of uncertainties in both modelling and observing, and are extremely challenging to model self-consistently since they involve physical processes that span across many orders of magnitude in physical scales.

The intrinsic production rate of ionizing photons, \Nion, is simply the number of ionizing photons being produced per second by a galaxy, which can be formally obtained by directly integrating over the galaxy SED
\begin{equation}
\dot{N}_\text{ion} = \int_{\nu_{912}}^{\infty} L_\nu(h\nu)^{-1}\; d\nu \text{.}
\label{eqn:xion}
\end{equation}
This quantity is often normalized to the intrinsic rest-frame far ultraviolet (FUV) luminosity and expressed as the ionizing photon production efficiency
\begin{equation}
\xi_\text{ion} = \dot{N}_\text{ion} \;/\; L_\text{FUV}\text{,}
\label{eqn:Nion}
\end{equation} 
where note that $L_\text{FUV}$ is the intrinsic UV luminosity, not including the effects of attenuation by dust, and is therefore in general not directly observable. The expected integrated value of \Nion\ or \xion\ for a galaxy therefore depends on its spectral energy distribution, which depends on the initial mass function (IMF) and joint distribution of ages and metallicities of the stellar populations contained in it, as well as on the physics of stellar evolution and stellar atmospheres.

Decades of effort have been invested into modelling the spectral energy distribution arising from `simple' (single age and metallicity) stellar populations (SPS models). Examples include \textsc{starburst99} \citep{Leitherer1999}, \textsc{P\'egase} \citep{Fioc1997, Fioc1999, Fioc2019}, \citet[hereafter \citetalias{Bruzual2003}, see also \citealt{Bruzual1983, Bruzual1993, Charlot1991}]{Bruzual2003}, \citet{Maraston2005}, Flexible Stellar Population Synthesis \citep[FSPS;][]{Conroy2009, Conroy2010, Conroy2010a}, Binary Population and Spectral Synthesis \citep[\textsc{bpass};][]{Stanway2016, Eldridge2017, Stanway2018}. These models combine stellar isochrones with stellar atmosphere models or templates, weight them with an assumed stellar initial mass function, and provide stellar continuum SEDs for a grid of stellar ages and metallicities. Although the predictions of SPS models are fairly robust and well-converged at optical wavelengths, the modelling of the massive stars that produce ionizing radiation suffers from much larger uncertainties, including the proper treatment of convection, rotation, and the evolution of binary stars. Moreover, it is unknown whether the stellar IMF during the EoR resembles the one in nearby galaxies, or whether the IMF has a significant dependence on the environment in which stars are born, which might lead to an indirect dependence on redshift or global galaxy properties.

In the past, coarse estimates of \xion\ have been obtained indirectly using the observed UV-continuum slope, $\beta_\text{UV}$, which is used as a proxy for predominant massive, UV-bright stars \citep{Robertson2013, Bouwens2015a, Bouwens2016b, Duncan2015} or using the measured stellar age and metallicities \citep[e.g.][]{Madau1999, Schaerer2003}. Recently, measurements of nebular emission lines from spectroscopy have enabled more direct constraints on \xion; both of which are used in conjunction with stellar population synthesis models. Recent studies have attempted to use measurement of nebular emission lines and UV-continuum fluxes to more directly infer the total number of LyC photons produced and put constraints on \xion. \citet{Bouwens2016a} and \citet{Lam2019} provided estimates with H$\alpha$ at $z\sim4$--5, and \citet{Stark2015} used \ion{C}{IV} as a tracer for a galaxy at $z \sim 7$. The highly-anticipated Near Infrared Spectrograph (NIRSpec) onboard the \textit{James Webb Space Telescope} (\textit{JWST}), as well as future Extremely Large Telescope (ELT) observations are expected to obtain high-resolution spectra for more robust determination of many physical properties including \xion. Moreover, low-redshift galaxies ($z \lesssim 2$) are sometimes viewed as analogues of their high-redshift counterparts, which may provide complementary insights into the otherwise hard-to-measure quantities \citep{Nakajima2016, Nakajima2018, Schaerer2016, Matthee2017, Shivaei2018}.

Many previous analytic calculations of reionization have assumed a constant value for \xion\ \citep[e.g.][]{Madau1999, Robertson2015} or adopted a simple parametrization as a function of redshift \citep[e.g.][]{Finkelstein2012a}. \citet{Kuhlen2012} attempted to explore the uncertainties in \xion\ due to lack of knowledge about the stellar populations and SED in their models by parametrizing it and exploring a bracketing range of values. \citet[hereafter \citetalias{Finkelstein2019}]{Finkelstein2019} parametrize \xion\ as a function of $M_\text{UV}$ and redshift and allow these parameters to be fit as part of a multi-parameter Bayesian Monte Carlo Maximum Likelihood fitting procedure. Many (semi-)numerical simulations that aim to capture the evolution of large-scale structure, but do not contain detailed modelling of galaxy formation, assume simple scaling relations that allow \xion\ or \Nion\ to scale with stellar mass or halo mass \citep{Mesinger2007, Choudhury2009, Choudhury2009a, Santos2010, Hassan2016, Hassan2017}.

\Nion\ and \xion\ for composite stellar populations depend on the joint distribution of ages and metallicities in each galaxy, and the values will be highly sensitive to the burstiness of the star formation history. These quantities may in principle evolve with redshift or be correlated with global galaxy properties such as mass or luminosity. It is thus important to use a self-consistent, physically grounded approach that hopefully contains the relevant internal correlations between physical properties. Cosmological models that follow the formation of stars and chemical evolution in galaxies can, in principle, straightforwardly obtain more physically motivated estimates of \xion\ by simply convolving their predicted star formation and chemical evolution histories with SPS models like those described above. \citet[hereafter \citetalias{Wilkins2016a}]{Wilkins2016a} did exactly this using the large volume \textsc{BlueTides} simulations, and presented predictions for how \xion\ evolves with redshift and its dependence on stellar mass. Similar work has also been done with the FIRE \citep{Ma2016a}, \textsc{Vulcan} \citep{Anderson2017}, and FirstLight zoom-in simulations \citep{Ceverino2019}.

In this paper, we focus on understanding the production efficiency of ionizing photons in EoR galaxies, which is a quantity that will be able to constrained with future \textit{JWST} photometric and spectroscopic surveys. Based on the results from the well-established physical models that have been rigorously tested in previous works, we explore the scaling relations between the intrinsic production efficiency of ionizing photons and a wide range of SF-related physical properties, as well as other observable properties. These results are also compared to predictions from other simulations and values inferred by existing observations. Our computationally efficient semi-analytic approach allows us to span a larger dynamic range in galaxy mass and in cosmic time than previous studies based on numerical hydrodynamical simulations. Note that only a fraction of the ionizing photons produced within the galaxy actually escape and make their way out into the intergalactic medium. Estimating this `escape fraction' is extremely difficult, and we do not attempt to address this issue in this paper.

In this series of \textit{Semi-analytic forecasts for JWST} papers, we make a wide variety of predictions for high-redshift galaxy populations that are expected to be detected by \textit{JWST} or other future instruments. We also investigate the implications of these galaxy populations for the reionization of the Universe. In \citet[hereafter \citetalias{Yung2019}]{Yung2019}, we presented distribution functions for the rest-frame UV luminosity and observed-frame IR magnitudes in \textit{JWST} NIRCam broadband filters. In \citet[hereafter \citetalias{Yung2019a}]{Yung2019a}, we further investigated the physical properties and the scaling relations for galaxies predicted by the same models. In this companion work (Paper III), we use these same models to make predictions for the intrinsic production rate of ionizing photons by star-forming galaxies. In the next paper \citet[hereafter \citetalias{Yung2020a}]{Yung2020a}, we combine our galaxy formation model with an analytic reionization model and a parametrized treatment of the escape fraction to create a physically motivated, source-driven pipeline to efficiently explore the effects of the predicted galaxy populations on cosmic reionization. All results presented in the paper series will be made available at \url{https://www.simonsfoundation.org/semi-analytic-forecasts-for-jwst/}. We plan on making full object catalogues available after the publication of the full series of papers.

The roadmap for this paper is as follows: the semi-analytic framework used in this work is summarized briefly in Section \ref{sec:sam}, where we also describe our procedure for calculating \Nion. We present our results in Section \ref{sec:results}, where we compare predictions across several different SPS models (Section \ref{sec:model_comp}), present the scaling relations and redshift evolution for these galaxy populations (Section \ref{sec:scaling}), and provide specific forecasts for future \textit{JWST} observations in (Section \ref{sec:scaling}). We then discuss our findings in Section \ref{sec:discussion}, and a summary and conclusions follow in Section \ref{sec:snc}.

\begin{figure*}
    \includegraphics[width=2\columnwidth]{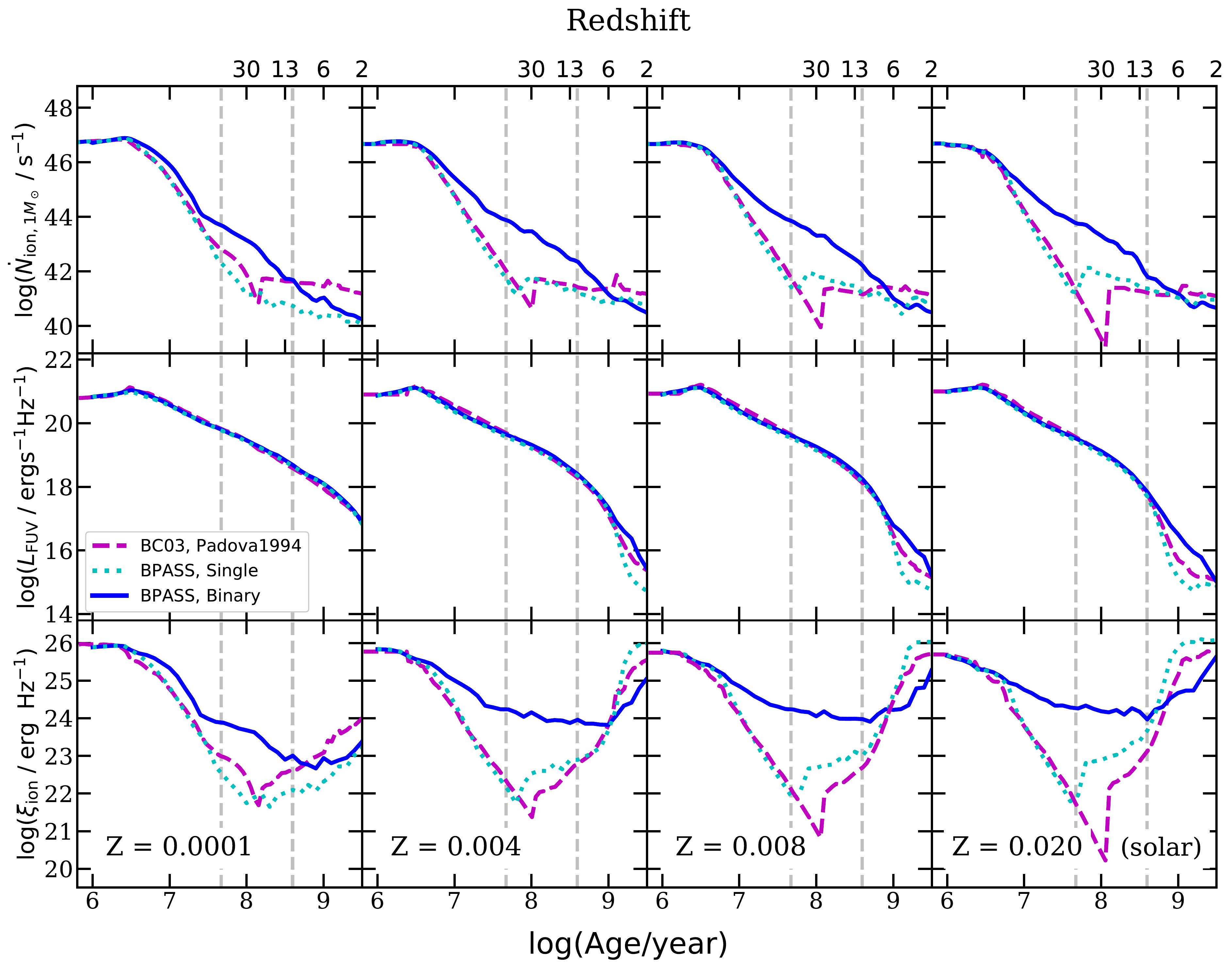}
    \caption{The evolution tracks of the production rate of ionizing photons per solar mass of stars formed ($\dot{N}_\text{ion,1\Msun}$; top row), the far-UV luminosity $L_\text{FUV}$ (also normalized to 1\Msun; middle row), and the production efficiency of ionizing photons (\xion; bottom row) as a function of the age of a stellar population, assuming an instantaneous starburst, predicted by the \citetalias{Bruzual2003} (purple dashed) and \textsc{bpass} single (cyan dotted) and binary (blue solid) SPS models. Each column shows a different metallicity as labelled. We also show the redshifts corresponding to the age equal to the age of the Universe, which can be taken as the upper limit on the stellar population age at that redshift. The vertical grey dashed lines show the age range that is most relevant to this work, where the upper (lower) bound is marked by the median of the mass-weighted stellar age for $z=4$ ($z=10$) galaxy populations predicted by our model.}
    \label{fig:age_stack}
\end{figure*}

\section{The semi-analytic framework}
\label{sec:sam}
The Santa Cruz semi-analytic model (SAM) for galaxy formation used in this study is very similar to the one outlined in \citet*[hereafter \citetalias{Somerville2015}]{Somerville2015}. The only changes are that we have implemented the updated \citet*{Okamoto2008} photoionization feedback recipe and updated the cosmological parameters to be consistent with the ones reported by the \citeauthor{Planck2016} in 2015: $\Omega_\text{m} = 0.308$, $\Omega_\Lambda = 0.692$, $H_0 = 67.8$\kms Mpc$^{-1}$, $\sigma_8 = 0.831$, and $n_s = 0.9665$. We recalibrated the model to $z\sim 0$ observations as described in \citetalias{Yung2019}, where all model components that are essential for this work, along with values of the free parameters, are also documented.
We refer the reader to the following works for full details of the modelling framework: \citet{Somerville1999}; \citet*{Somerville2001}; \citet[hereafter \citetalias{Somerville2008}]{Somerville2008}; \citet[hereafter \citetalias{Somerville2012}]{Somerville2012}; \citet*[hereafter \citetalias{Popping2014}]{Popping2014} and \citetalias{Somerville2015}. Throughout this work, we adopt the cosmological parameters specified above.

The semi-analytic approach to modelling galaxy formation relies on the merger histories of dark matter halos, more commonly known as `merger trees', which can either be extracted from cosmological $N$-body simulations or constructed semi-analytically based on the Extended Press-Schechter (EPS) formalism \citep{Press1974, Lacey1993}. The EPS-based method, which has been shown to be in good qualitative agreement with $N$-body simulations \citep{Somerville1999a, Somerville2008, Zhang2008, Jiang2014}, is able to efficiently resolve merger histories with very high mass resolution at very low computational cost. We are able to use this method to sample halos over an extremely wide dynamic range, from the ones near the atomic cooling limit to the most massive halos that have collapsed at a given redshift. At each output redshift, we set-up a grid of root halos spanning the range in virial velocity $V_\text{vir} \approx 20$--500 \kms. And for each root halo in the grid, one hundred Monte Carlo realizations of the merger histories are generated, each traced down to progenitors of a minimum resolution mass of either $M_\text{res}\sim10^{10}$\Msun\ or 1/100th of the root halo mass, whichever is smaller. The expected volume-averaged abundances of these halos are assigned based on the halo mass function (HMF) from the Bolshoi-Planck simulation from the MultiDark suite \citep{Klypin2016} with fitting functions provided in \citet{Rodriguez-Puebla2016}, which has been examined up to $z \approx 10$.

In the latest version of the model \citepalias{Popping2014, Somerville2015}, recipes for multiphase gas-partitioning and \molh-based star formation have been implemented and tested. In the former, the disc component of each galaxy is divided into annuli and the cold gas content in each annulus is partitioned into an atomic (\ion{H}{I}), ionized (\ion{H}{II}), and molecular (\molh) component. In the latter, the surface density of SFR ($\Sigma_\text{SFR}$) is estimated based on the surface density of molecular hydrogen ($\Sigma_\text{\molh}$) using observationally motivated, empirical \molh-based SF relations. \citetalias{Somerville2015} investigated several different gas partitioning and SF recipes, and favoured the metallicity-based, UV-background-dependent gas partitioning recipe, which is based on simulations by \citet[hereafter \citetalias{Gnedin2011}]{Gnedin2011}. In the star formation recipe, $\Sigma_\text{SFR}$ is expressed as a broken power law function of $\Sigma_\text{\molh}$. \citetalias{Somerville2015} investigated a power law with a single slope, and one in which the slope becomes steeper above a critical \molh\ surface density (`two-slope'). They found that both recipes produce similar results at $z\sim 0$, but at high redshift ($z\sim 4$--6), the `two-slope' recipe was favoured, in agreement with direct evidence from recent observations \citep[e.g.][]{Sharon2013, Rawle2014, Hodge2015, Tacconi2018}. In \citetalias{Yung2019} and \citetalias{Yung2019a}, we investigated both star formation laws up to $z\sim 10$, and concluded even more strongly that the `two-slope' relation is much more consistent with observations of luminous high redshift galaxies. In this paper and the remainder of the \textit{series}, we therefore adopt the `two-slope' star formation relation, which we have referred to as the `fiducial' \citetalias{Gnedin2011}-\citetalias{Bigiel2008}2 model throughout this paper series.

Our model consists of a collection of physically motivated, phenomenological and empirical recipes, including cosmological accretion and cooling, stellar-driven winds, chemical evolution, black hole growth and feedback, galaxy mergers, etc. These recipes are then used to track the evolution of a wide range of global physical properties of galaxies, such as stellar mass, star formation rate, masses of multiple species of gas (ionized, atomic, molecular), and stellar and gas phase metallicity, etc. For each galaxy, synthetic spectral energy distributions (SEDs) are created by convolving the predicted stellar age and metallicity with stellar population synthesis (SPS) models of \citet[hereafter \citetalias{Bruzual2003}]{Bruzual2003} with the Padova1994 \citep{Bertelli1994} isochrones assuming a universal Chabrier stellar initial mass function \citep[IMF,][]{Chabrier2003a}. Throughout this work, all magnitudes are expressed in the AB system \citep{Oke1983} and all uses of log are base 10 unless otherwise specified.

\begin{figure*}
    \includegraphics[width=2\columnwidth]{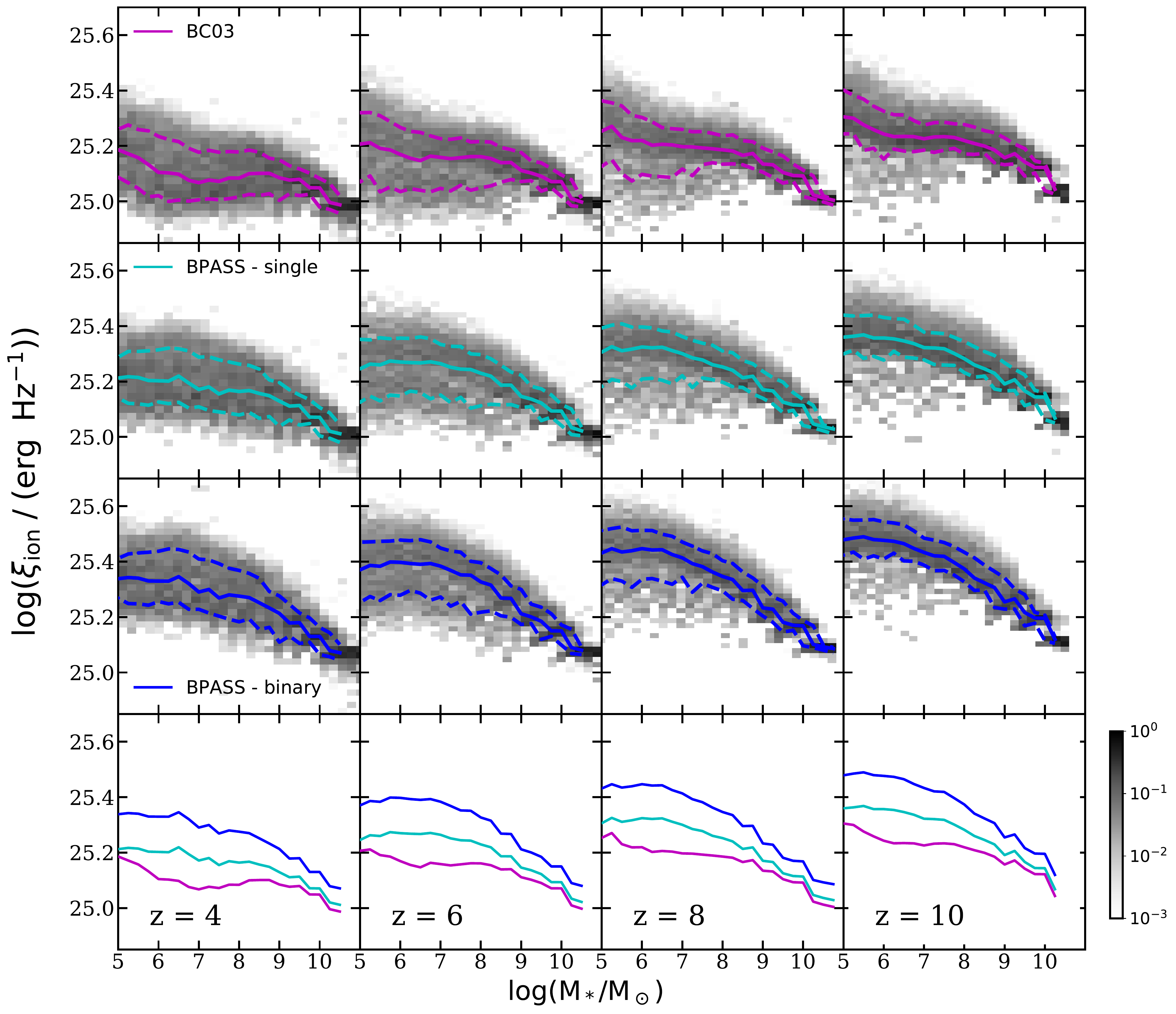}
    \caption{Conditional distributions of \xion\ versus stellar mass predicted at $z=4$, 6, 8, and 10 using SPS models from \citetalias{Bruzual2003} (purple) and \textsc{bpass} (cyan and blue for single and binary). Each column shows a different redshift as labelled. The grayscale 2D histograms show the conditional number density of galaxies in each bin, normalized to the sum of the number density per Mpc$^{3}$ in each corresponding (vertical) stellar mass bin.The blue solid and dashed lines show the 50th, 16th and 84th percentiles, respectively, for the distribution in each panel. The medians of these distributions are repeated in the bottom row for comparison, using the same colour coding as in the other panels.}
    \label{fig:xion_mstar_allmodel} 
\end{figure*}

\subsection{Calculating \Nion}
\label{sec:Nion}
As a useful reference, in fig. \ref{fig:age_stack} we first show the dependence of \Nion, $L_{\rm UV}$, and \xion\ as a function of age for simple stellar populations, assuming an instantaneous starburst, of different metallicities ($Z = 0.0001$, 0.004, 0.008, and 0.020), in two popular sets of SPS models, \citetalias{Bruzual2003} and \textsc{bpass}. We show the \textsc{bpass} models that account only for single star evolution, as is also the case in the \citetalias{Bruzual2003} models, and those that account for binary star evolution. These quantities are directly obtained from the data tables released by the \citetalias{Bruzual2003}\footnote{\url{http://www.bruzual.org/~gbruzual/bc03/}} and \textsc{bpass} groups\footnote{\url{https://bpass.auckland.ac.nz/}, v2.2.1} \citep{Stanway2016, Eldridge2017, Stanway2018}. Both SPS models adopted in this work assume a Chabrier IMF with an upper mass cutoff $m_\text{U} = 100$\Msun. In the top panel of fig. \ref{fig:age_stack}, we show the evolution of ionizing photon production rate normalized to that for a single solar mass of stars, $\dot{N}_\text{ion,1\Msun}$. As discussed by \citet{Stanway2016}, this quantity is strongly affected by both binary evolution and metallicity. However, in the middle panel of fig. \ref{fig:age_stack} we see that differences in rest-frame far-ultraviolet (FUV) luminosity, $L_\text{FUV}$, predicted by different models are rather small for young stellar populations. As a result, in the bottom panel, which shows \xion, we can see that the value of this quantity can differ by as much as several orders of magnitude at intermediate ages for models that include binary evolution relative to those that do not.

In this work, we convolve these SPS predictions for \Nion\ with the predicted 2D histogram of stellar ages and metallicities present in each galaxy at a given output redshift. This histograms are built up as follows. At each time step during each galaxy's evolution, a `star parcel' d$m_*$ is created with a mass determined by the \molh\ density in the disc and the \molh-based SF recipe. Each star parcel inherits the metallicity of the cold gas in the ISM at the time it forms. Metals are deposited in the cold gas by new stars with an assumed yield, and may be removed from the ISM by stellar driven winds (see \citetalias{Somerville2015} for details). When galaxies merge, their stellar populations are combined. At an output redshift, we evaluate \Nion\ for each galaxy by summing up \Nion\ for each star parcel from the data tables provided by a chosen SPS model based on its stellar age and metallicity at the time.

In the previous papers in the series, we adopted the \citetalias{Bruzual2003} SPS models to compute the galaxy SEDs, and the model UV luminosity functions (UV LFs) have been extensively tested against existing observations. In order to avoid repeating these comparisons, we retain the estimates of $L_\text{FUV}$ based on the \citetalias{Bruzual2003} models in all of our results, and only compute \Nion\ with the SPS model variants. Thus the values of \xion\ for the \textsc{bpass} models are not strictly self-consistent. However, we saw in fig. \ref{fig:age_stack} that the predictions for $L_\text{FUV}$ from these two SPS models are very similar for the ages that are most relevant for our study ($z \gtrsim 6$), so this should not cause a significant discrepancy. Furthermore, \xion\ is only used for illustrative purposes. For actually computing the reionization history, we use \Nion\ directly.

\begin{figure*}
    \includegraphics[width=1.8\columnwidth]{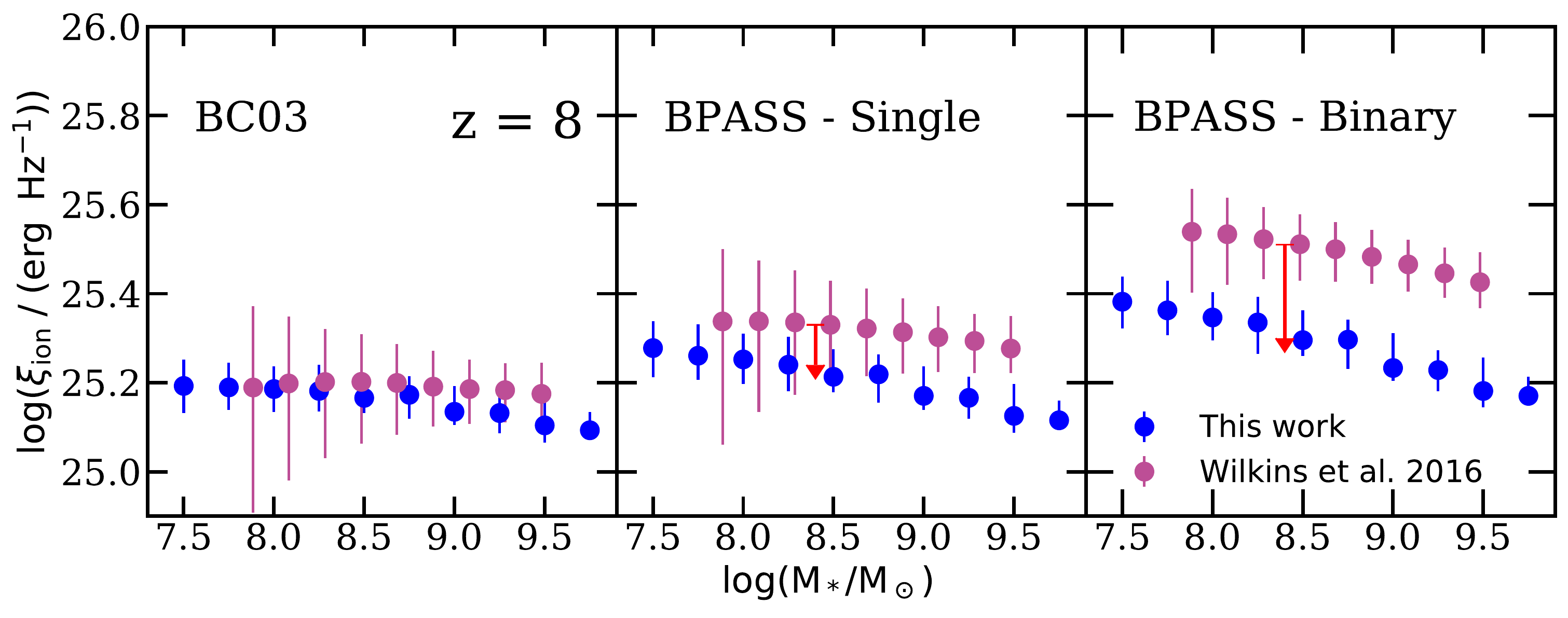}
    \caption{A comparison of \xion\ and \Nion\ predicted with \citetalias{Bruzual2003} and \textsc{bpass} (both single and binary) SPS models between this work and \citetalias{Wilkins2016a}. The data points mark the median and the error bars span the 16th to 84th percentile range. The red arrow shows the estimated difference in \xion\ changing from \textsc{bpass} v2.0 to v2.1 models, accounting for the metallicities difference between the SAM and \textsc{BlueTides} galaxies, see text for details.}
    \label{fig:comparew16_xion} 
\end{figure*}

\begin{figure}
    \includegraphics[width=\columnwidth]{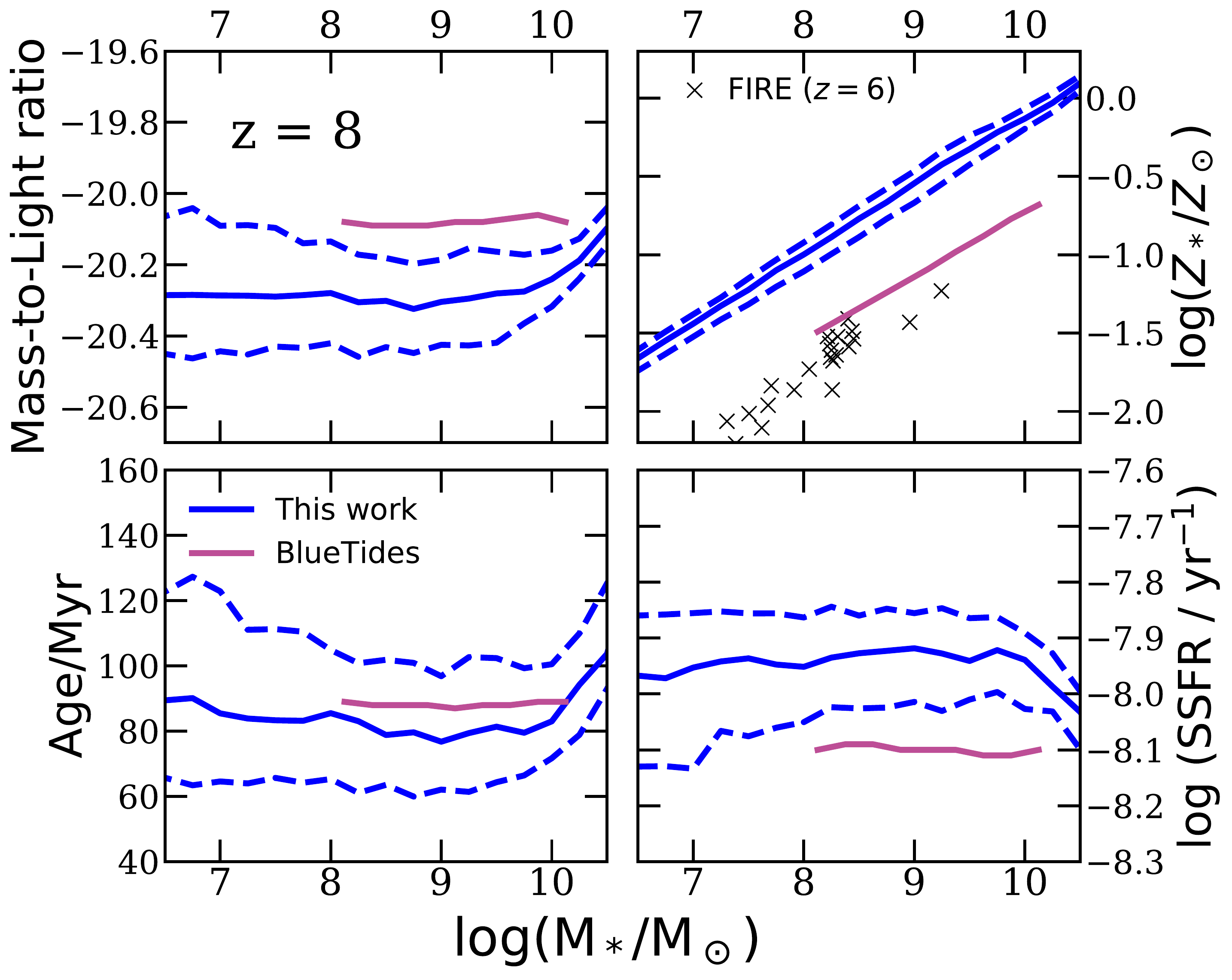}
    \caption{Comparison between the median of the intrinsic rest-UV mass-to-light ratio ($\log(( M_*/{\rm M}_\odot)/(L_{\nu, {\rm FUV}}/{\rm erg\;s}^{-1}{\rm Hz}^{-1}))$; upper left), stellar metallicity (upper right), mass-weighted stellar age (lower left), and specific SFR (lower right) for galaxies predicted by our SAM (blue) and \textsc{BlueTides} \citep[purple]{Wilkins2017}. The black crosses show the stellar metallicity predicted by the FIRE simulations at $z = 6$ \citep{Ma2016}. The dashed blue lines mark the 16th and 84th percentiles of the SAM predicted distributions. }
    \label{fig:comparew16_prop} 
\end{figure}

\section{Results}
\label{sec:results}
Our results are organized in three parts. 1) We compare results across different SPS models. 2) We present scaling relations and redshift evolution for ionizing photon production rates and efficiencies, and show how \xion\ relates to observable properties. 3) We show how variations in the SN feedback model impact our predictions for galaxies during the EoR.

\subsection{Comparison across SPS models}
\label{sec:model_comp}
In this subsection, we compare \xion\ computed with several SPS models, while other model components in our SAM remain in their fiducial configurations. As illustrated in fig. \ref{fig:age_stack}, the predicted production rate of ionizing photons depends on the age and metallicity of the stellar populations, and these relations vary across different SPS models. Note that these SPS models are applied to the same galaxy populations. In other words, the galaxies used for all three SPS models have identical star formation histories and physical properties.

In fig. \ref{fig:xion_mstar_allmodel}, we show the distributions of \xion\ against $M_*$ for \citetalias{Bruzual2003} and \textsc{bpass} models for single and binary stellar population at selected redshifts. The 2D histograms illustrate the conditional number density per Mpc$^{3}$ of galaxies in each bin and are normalized to the sum of the number density in the corresponding (vertical) stellar mass bin. The 16th, 50th, and 84th percentiles are marked in each panel to illustrate the statistical distribution. The median of predictions of all models for each given redshift are overlaid for easy comparison.

In general, we see that massive galaxies have lower \xion\ than low-mass galaxies regardless of the choice of SPS models and redshift. Also the scatter of the relation is larger for low-mass galaxies. Although all models demonstrate some level of dependence on $M_*$, we find that \citetalias{Bruzual2003} yields the weakest dependence for $M_* < 10^8$ \Msun, but both \textsc{bpass} models predict more evolution for galaxies in the same mass range. The differences between predictions from the most and the least optimistic SPS model can be up to $\sim0.2$ dex for galaxies with $M_*\sim10^7$\Msun, and the difference shrinks for more massive galaxies.

At fixed $M_*$, the predicted value of \xion\ evolves mildly as a function of redshift, with an average downward shift of $\sim 0.1$ dex from $z=10$ to $z=4$. High redshift galaxies seem to be more efficient at producing ionizing photons than their low-redshift counterparts of similar mass due to the younger stellar age and lower metallicity. We also notice the slope of the \xion-$M_*$ relation becomes more shallow with decreasing redshift, likely due to the corresponding change in the slope of the $M_*$-$Z_*$ relation.

\begin{figure*}
    \includegraphics[width=2\columnwidth]{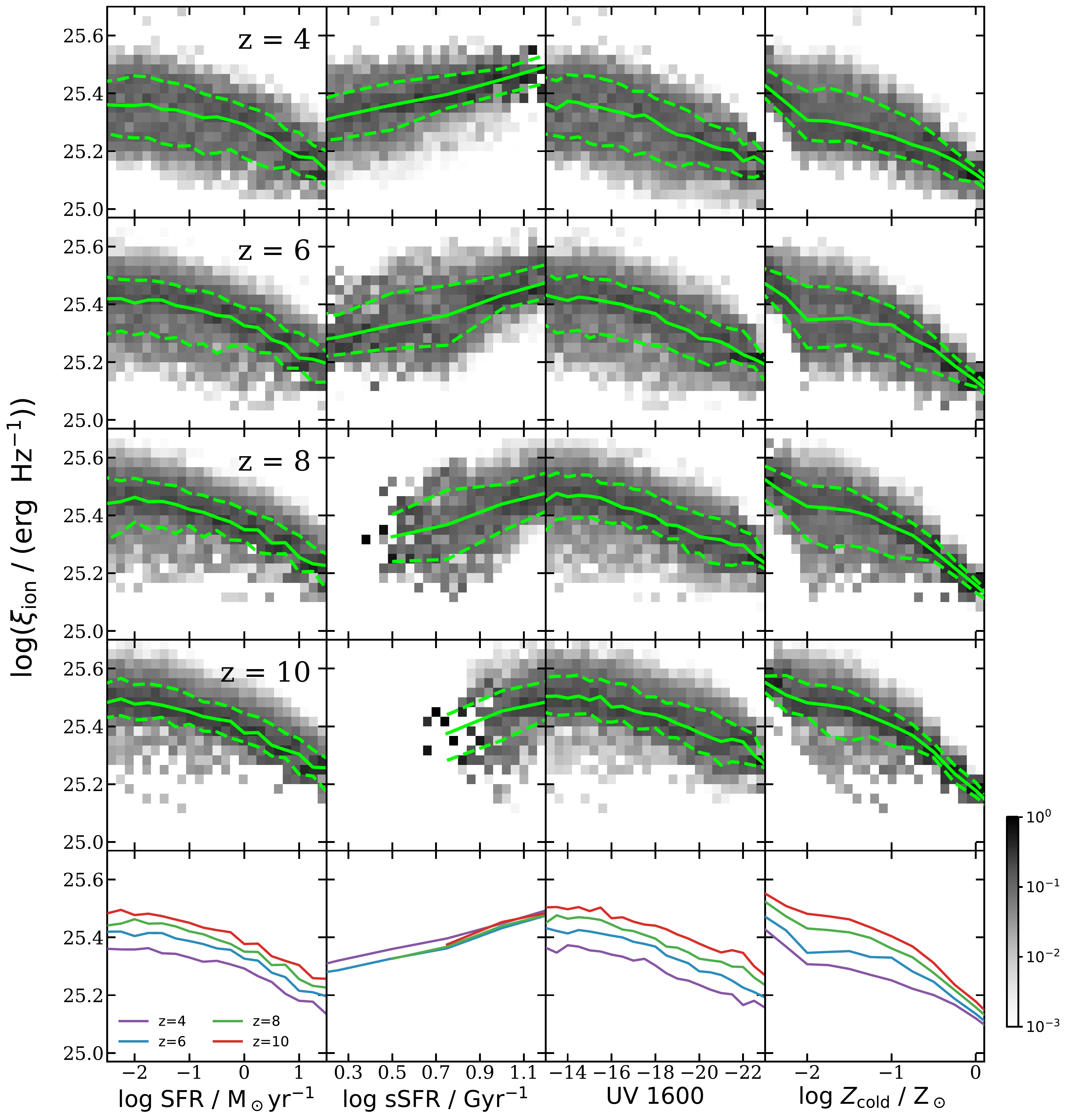}
    \caption{Conditional distributions of \xion\ versus SFR (averaged over 100 Myr), specific SFR (sSFR), intrinsic UV luminosity, and cold gas metallicity at $z=4$, 6, 8, and 10 with our fiducial model configurations with the \textsc{bpass} binary SPS models. The green solid and dashed lines mark the 50th, 16th, and 84th percentiles. The greyscale 2D histograms show the conditional number density of galaxies in each bin, normalized to the sum of the number density per Mpc$^{3}$ in the corresponding (vertical) bin. The median of these distributions are repeated in the last row for comparison to illustrate the redshift evolution.}
    \label{fig:prop_xion_fidmodel} 
\end{figure*}

\subsection{Comparison with numerical hydrodynamic simulations}

We compare our results with predictions of \xion\ computed from galaxies in the \textsc{BlueTides} simulations with various SPS models presented in \citetalias{Wilkins2016a}. \textsc{BlueTides} is a cosmological hydrodynamical simulation with 400 Mpc $h^{-1}$ on a side and $2\times7040^3$ particles, which is able to resolve galaxies with $M_* \gtrsim 10^8$ providing a wide variety of predictions for physical and observable properties for high-redshift galaxies up to $z \sim 8$ \citep{Feng2016, Wilkins2017}. We also note that \textsc{BlueTides} adopted a set of cosmological parameters that are consistent with constraints from WMAP9. In fig. \ref{fig:comparew16_xion}, we compare the \xion\ relations at $z = 8$ predicted by our SAM and \textsc{BlueTides} with \citetalias{Bruzual2003} and \textsc{bpass} single and binary star models \citep[v2.0,][]{Stanway2016}, which does not include nebular emission. The \citetalias{Wilkins2016a} calculations assumed a \citet{Salpeter1955} IMF. We convert the quoted stellar masses to a Chabrier IMF by adding -0.21 dex for direct comparison \citep{Madau2014}. The values of \xion\ are nearly unaffected by this change in IMF.
We note that the predictions for \xion\ from \citetalias{Wilkins2016a} and our models are very consistent for the
\citetalias{Bruzual2003} models, while the SAM predictions are about 0.1 dex lower for the \textsc{bpass} single and about 0.25 dex lower for the \textsc{bpass} binary models (though the qualitative trends are the same).

In order to better understand these differences, we further
compare a number of key physical properties for our model galaxies to the predictions by \textsc{BlueTides} reported in \citet{Wilkins2017} in fig. \ref{fig:comparew16_prop}, including intrinsic (dust-free) UV mass-to-light ratio, stellar metallicity, mass-weighted stellar age, and specific star formation rate (sSFR). Note that this work assumed a Chabrier IMF, which is consistent with our work and does not require any adjustment.
We also compare to the mass-metallicity relation predicted by the FIRE (Feedback in Realistic Environment) simulations \citep{Hopkins2014, Ma2016}, which are high-resolution cosmological hydrodynamic zoom-in simulations. We show predictions at $z=6$ in our comparison, which is the highest redshift provided in the FIRE publications. The FIRE simulations assumed a \citet{Kroupa2001} IMF and we convert their stellar masses to a Chabrier IMF by adding -0.03 dex \citet{Madau2014}. It is interesting that although \textsc{BlueTides} and FIRE have very different resolution as well as different subgrid recipes for star formation and stellar feedback, the predicted mass-metallicity relations are very similar, and much lower than the one predicted by our SAM. We discuss possible reasons for this discrepancy in Section~\ref{sec:discussion}, however, we do not currently have a complete understanding of this issue, and it is beyond the scope of this paper to determine.

The differences in the stellar population properties between the SAM and \textsc{BlueTides} galaxies do not seem sufficient to explain the difference in the predicted value of \xion, nor it is obvious how this would explain why the differences are more pronounced for the \textsc{bpass} models. However, \citetalias{Wilkins2016a} used the \textsc{bpass} v2.0 version of the models, while we use the latest v2.2.1 version. \citet{Eldridge2017} provide a detailed discussion of the implications of changes from the v2.0 to v2.1 models for the production efficiency of ionizing photons (see their Section 6.6.1). They find that the use of improved stellar atmosphere models for low-metallicity O stars leads to a less dramatic increase in \xion\ from the single star to binary star case than was seen in the v2.0 models, as well as a weaker dependence of \xion\ on metallicity. We make use of their figure 35 to estimate the effect of changing from the v2.0 to v2.1 models, assuming the typical metallicity of the SAM and \textsc{BlueTides} galaxies. This correction is shown by the red arrow in our fig. \ref{fig:comparew16_xion}, showing that this largely accounts for the discrepancy.

\begin{figure}
    \includegraphics[width=\columnwidth]{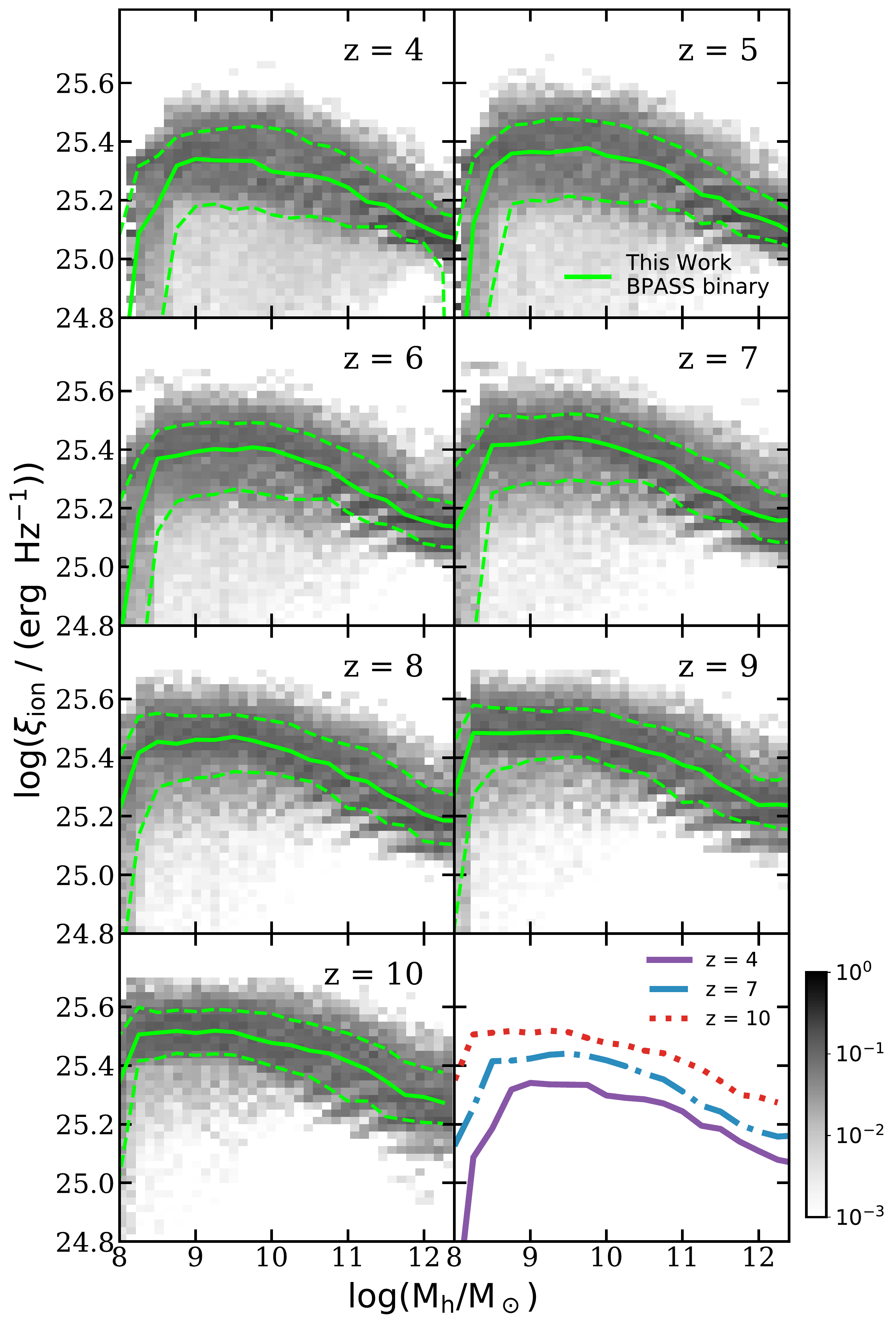}
    \caption{Predicted \xion\ as a function of halo mass between $z=4$--10 predicted with the \textsc{bpass} binary model. The green solid and dashed lines mark the 50th, 16th, and 84th percentiles. The 2D histograms are colour-coded to show the conditional number density per Mpc$^3$ of galaxies in each bin, normalized to the sum of the number density in the corresponding (vertical) halo mass bin. The last panel show an overlay of the SHMR median predicted at $z = 4$, 7, and 10 from our model.}
    \label{fig:xionMH_bin}
\end{figure}

\begin{figure}
    \includegraphics[width=\columnwidth]{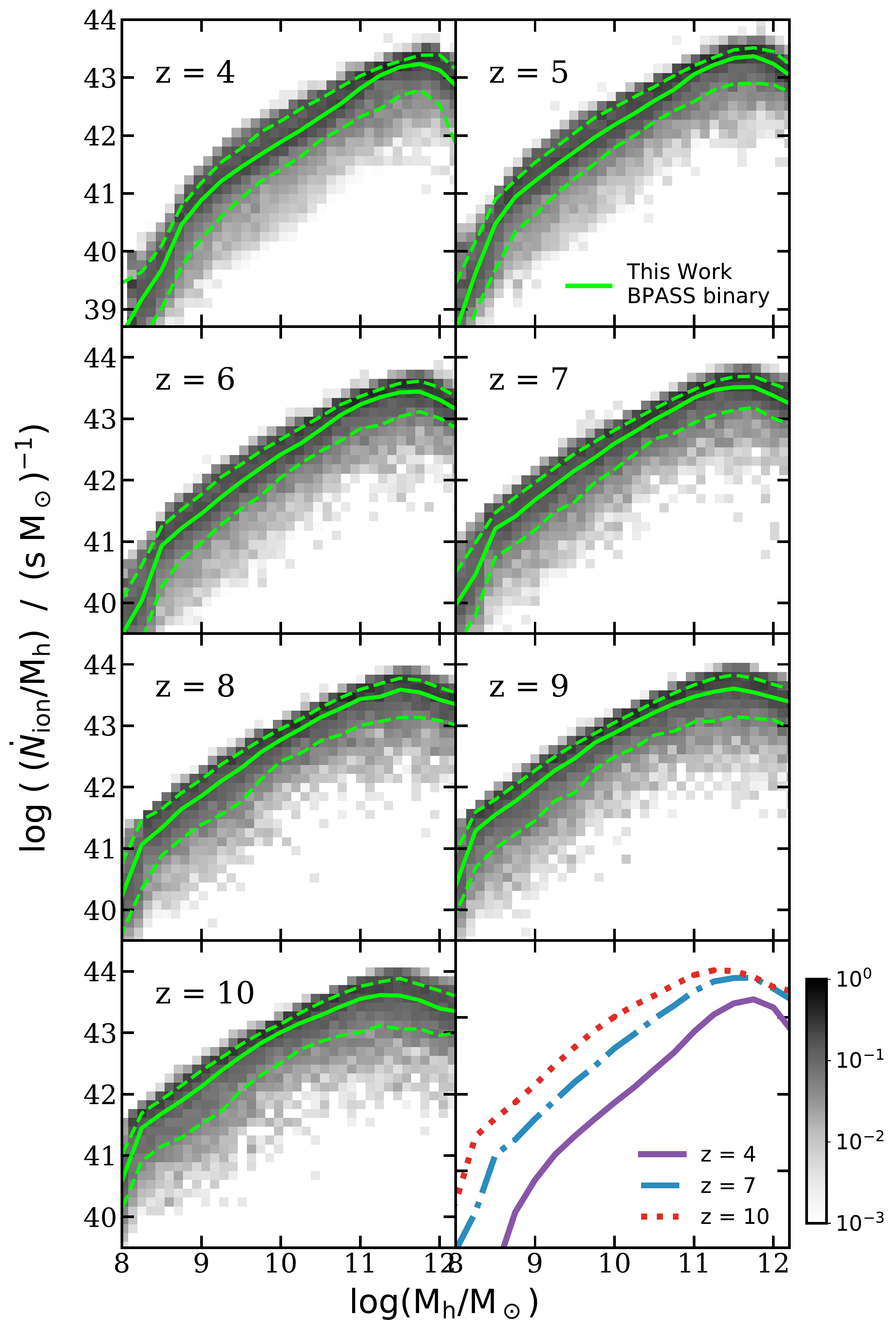}
    \caption{Specific ionizing photon production rate, $\dot{N}_\text{ion}/M_\text{h}$, as a function of halo mass between $z=4$--10 predicted with the \textsc{bpass} binary model. The green solid and dashed lines mark the 50th, 16th, and 84th percentiles. The greyscale 2D histograms show the conditional number density per Mpc$^3$ of galaxies in each bin, normalized to the sum of the number density in the corresponding (vertical) halo mass bin. The last panel shows the median $\dot{N}_\text{ion}/M_\text{h}$ relations predicted at $z = 4$, 7, and 10 from our model.}
    \label{fig:NionMH_bin}
\end{figure}

\begin{figure*}
    \includegraphics[width=2\columnwidth]{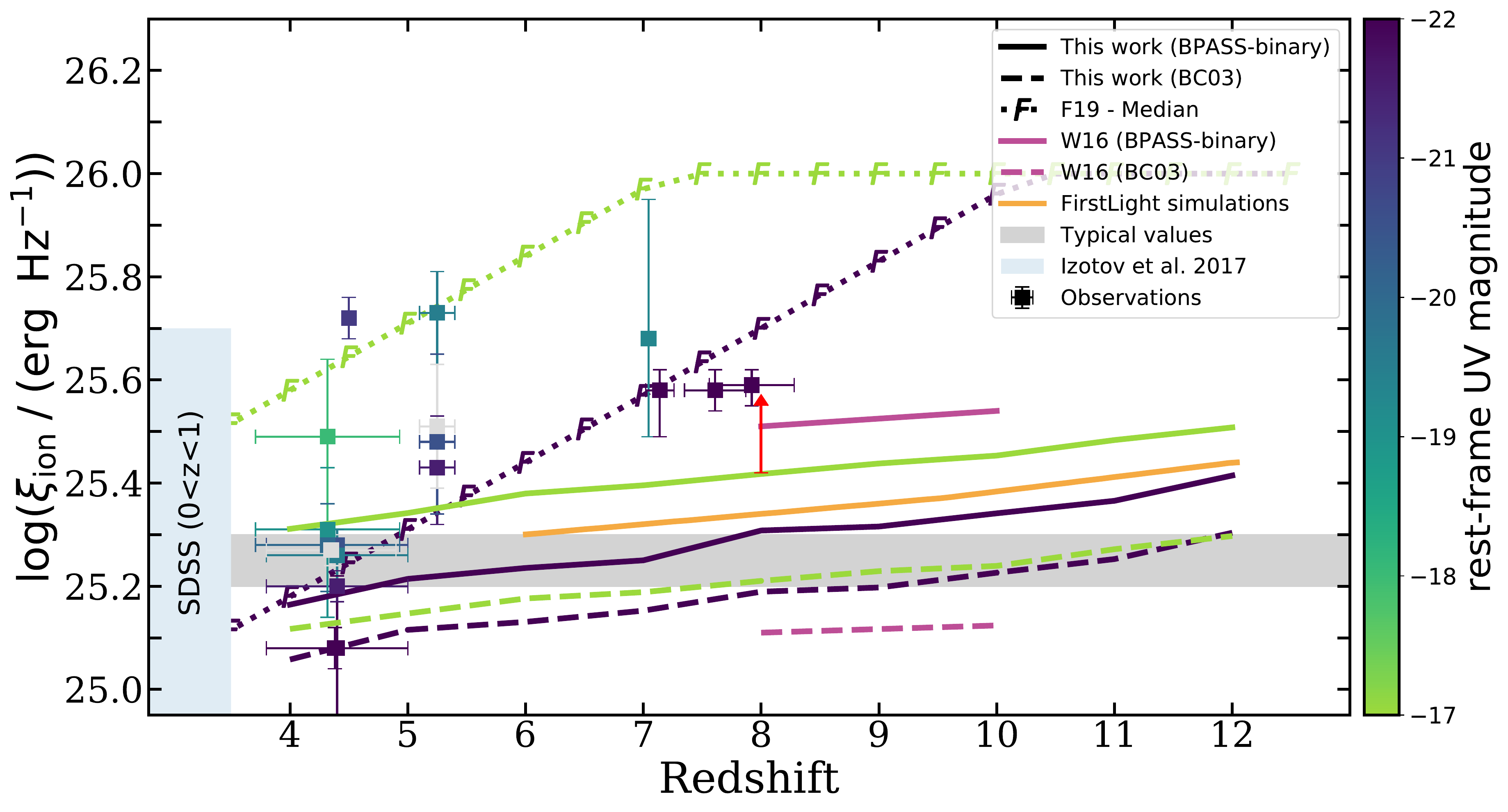}
    \caption{A comparison of the redshift evolution of \xion\ predicted by our models to a compilation of observations and empirical constraints. The solid and dashed lines denote our predictions using \citetalias{Bruzual2003} (solid) and the \textsc{bpass} binary SPS (dashed) models and are colour-coded for rest-frame dust-attenuated $M_\text{UV}$. Here we show the two bracketing cases representing bright ($M_\text{UV} = -22$) and faint ($M_\text{UV} \sim =-17$) galaxies shown in dark blue and green, respectively. The red arrow shows an estimate of the increase in \xion\ if the upper mass limit of the IMF were changed from 100 \Msun\ to 300 \Msun. We also plot the median parameters obtained by \citetalias{Finkelstein2019} for the same bracketing $M_\text{UV}$, shown with a dotted line with a `\textit{F}' marker. The solid and dashed purple lines show the luminosity-weighted value of \xion\ for galaxies with $M_* > 10^8$ \Msun\ at $z=8$--10 predicted by \citetalias{Wilkins2016a} with the SPS models considered in this work. The solid orange line shows results from the FirstLight simulations, which assumed a \textsc{bpass} binary SPS model \citep{Ceverino2019}. We include a compilation of observations from \citet{Shim2011, Stark2015, Stark2017, Bouwens2016a, Smit2016, Rasappu2016, Lam2019} for comparison; see text for a description. All observations and model predictions are colour-coded for rest-frame UV magnitude. Typical values adopted in analytic calculations are highlighted by the grey band. We also show the range derived from a large sample of SDSS galaxies at $0 < z < 1$ \citep{Izotov2017}. The lower bound of this range extends to $\xi_\text{ion} = 24$ (not shown).}
    \label{fig:xion_redshift} 
\end{figure*}

\subsection{Scaling relations for the production rate and efficiency of ionizing radiation}
\label{sec:scaling}

In this subsection, we present scaling relations for \xion\ and \Nion\ with respect to selected observable and physical properties predicted by our fiducial model configurations with the \textsc{bpass} binary SPS models. We also study the redshift evolution of these relations and compare to recent observations.

We show the distribution of \xion\ against SFR (averaged over 100 Myr), specific SFR (sSFR), intrinsic rest-frame $M_\text{UV}$ (without attenuation by dust), and cold gas metallicity ($Z_\text{cold}$) predicted at $z = 4$, 6, 8, and 10 in fig. \ref{fig:prop_xion_fidmodel}. Throughout this work, we refer to the UV luminosity without a correction for attenuation effect by dust as the intrinsic $M_\text{UV}$, which is to distinguish it from the dust-attenuated $M_\text{UV}$. Similar to plots in previous sections, the 2D histograms indicate the conditional number density per Mpc$^{3}$ of galaxies in each bin and is normalized to the sum of the number density in its corresponding (vertical) stellar mass bin. The 16th, 50th, and 84th percentiles are marked in each panel. The median of predictions made with all models for each given redshift are overlaid to illustrate the redshift evolution.

We find mild trends of decreasing \xion\ with increasing SFR, increasing \xion\ with increasing sSFR, decreasing \xion\ with increasing UV luminosity, and decreasing \xion\ with increasing gas phase metallicity. The inverse trends of \xion\ with SFR and UV luminosity may appear counter-intuitive, until we recall that both of these quantities are strongly correlated with stellar mass, and hence with metallicity, so the inverse correlation between metallicity and \xion\ is likely driving this result. As seen in fig. \ref{fig:comparew16_prop}, there is only a weak trend between stellar mass and mean stellar age in our models. At fixed SFR, UV luminosity, and cold gas metallicity, \xion\ decreases with cosmic time from $z\sim 10$ to 4, due to the increasing fraction of relatively old stars. Only for fixed sSFR is \xion\ nearly constant with redshift, indicating that this quantity is correlated with the fraction of young stars.

Fig. \ref{fig:xionMH_bin} shows the distribution of \xion, predicted for halo populations between $z=4$--10. We see that for halos of the same mass, the production efficiency of ionizing photons can change by up to $\sim0.2$ dex between $z=10$ and 4. Similarly, fig. \ref{fig:NionMH_bin} shows the distribution of the specific ionizing photon production rate, $\dot{N}_\text{ion}/M_\text{h}$, predicted for the same set of halos, where the production rate of ionizing photons can be up to $\sim1.5$ orders of magnitude higher at $z=10$ relative to $z=4$ for halos of the same mass. This relation evolves much more rapidly than the stellar-to-halo mass ratio (SHMR) presented in \citetalias{Yung2019a}. This is due to our finding (presented in \citetalias{Yung2019a}) that galaxies at high redshift are intrinsically brighter in the UV than their low-redshift counterparts of similar stellar mass due to the higher SFR, younger stellar population, and lower metallicities.

On the other hand, \xion\ does not evolve as much across redshift since both \Nion\ and $L_\text{FUV}$ evolve in the same direction. In fig. \ref{fig:xion_redshift}, we illustrate the evolution of \xion\ as a function of redshift. We show predictions from our fiducial model with both \textsc{bpass} and \citetalias{Bruzual2003} SPS models at the bracketing UV magnitudes of $M_\text{UV} = -22$ and $-17$ and compare these to a compilation of observational estimates. Both our predictions and the observations are colour-coded for rest-frame $M_\text{UV}$. These measurements are derived from H$\alpha$ flux and UV-continuum luminosity \citep{Bouwens2016a, Smit2016, Rasappu2016, Lam2019}, H$\alpha$ flux and UV-based SFR \citep{Shim2011, Marmol-Queralto2016}, \ion{C}{IV} measurement \citep{Stark2015}, and SED fitting \citep{Stark2017}. We indicate the range of $\log\xi_\text{ion} = 24.0$ to 25.7 derived from $\sim$14,000 compact star-forming galaxies found in the Sloan Digital Sky Survey (SDSS; DR12) between $0 < z < 1$ for comparison \citep{Izotov2017}. We also include the luminosity-weighted \xion\ from \citetalias{Wilkins2016a} at $ = 8$--10. We compare our predictions to the results from the FirstLight simulations, which is a set of high-resolution zoom-in simulations based on cosmological boxes of 10, 20, and 40 Mpc h$^{-1}$ on a side, resolving galaxies in halos with circular velocities spanning the range between 50--250 km s$^{-1}$ at $z\gtrsim5$ \citep{Ceverino2017, Ceverino2019}. The computation for \xion\ is based on synthetic SEDs from \textsc{bpass} \citep[v2.1,][]{Eldridge2017} including nebular emission \citep*{Xiao2018} and a Kroupa-like IMF. Note that whilst the simulations have been run with both WMAP and Planck cosmologies, the results referenced here are only available with the WMAP-5 cosmology. For reference, we also label the range of typical values $\xi_\text{ion} \approx 25.20$--25.30 that are assumed in many analytic studies \citep[e.g.][]{Bouwens2012, Finkelstein2012a, Kuhlen2012, Duncan2015, Robertson2015}. Our models predict very mild evolution in \xion\ as a function of redshift and a very consistent $\sim0.1$--0.15 dex difference between the bright ($M_\text{UV} = -22$) and faint ($M_\text{UV} = -17$) galaxies. The red arrow shows an estimate (from fig. 35 of \citet{Eldridge2017}) of the increase in \xion\ if the upper mass limit of the IMF were changed from 100 \Msun\ to 300 \Msun, illustrating how sensitive \xion\ is to the upper end of the stellar IMF. In addition, lower-redshift measurements reporting $\xi_\text{ion} \gtrsim 25.5$ at $z\sim3$ appear to be in tension with standard model assumptions \citep{Nakajima2016, Nakajima2018}.
We also note that estimates of \xion\ from detected rest-UV emission lines at higher redshifts (e.g. $z>5$) tend to be obtained for sources with detectable nebular lines. As there is a general correlation between the equivalent width (EW) in these lines and \xion, the published \xion\ values based on rest-UV lines are likely biased towards the upper end of the true range of values, rather than reflecting the mean or median of the overall population at these epochs.

\begin{figure}
    \includegraphics[width=\columnwidth]{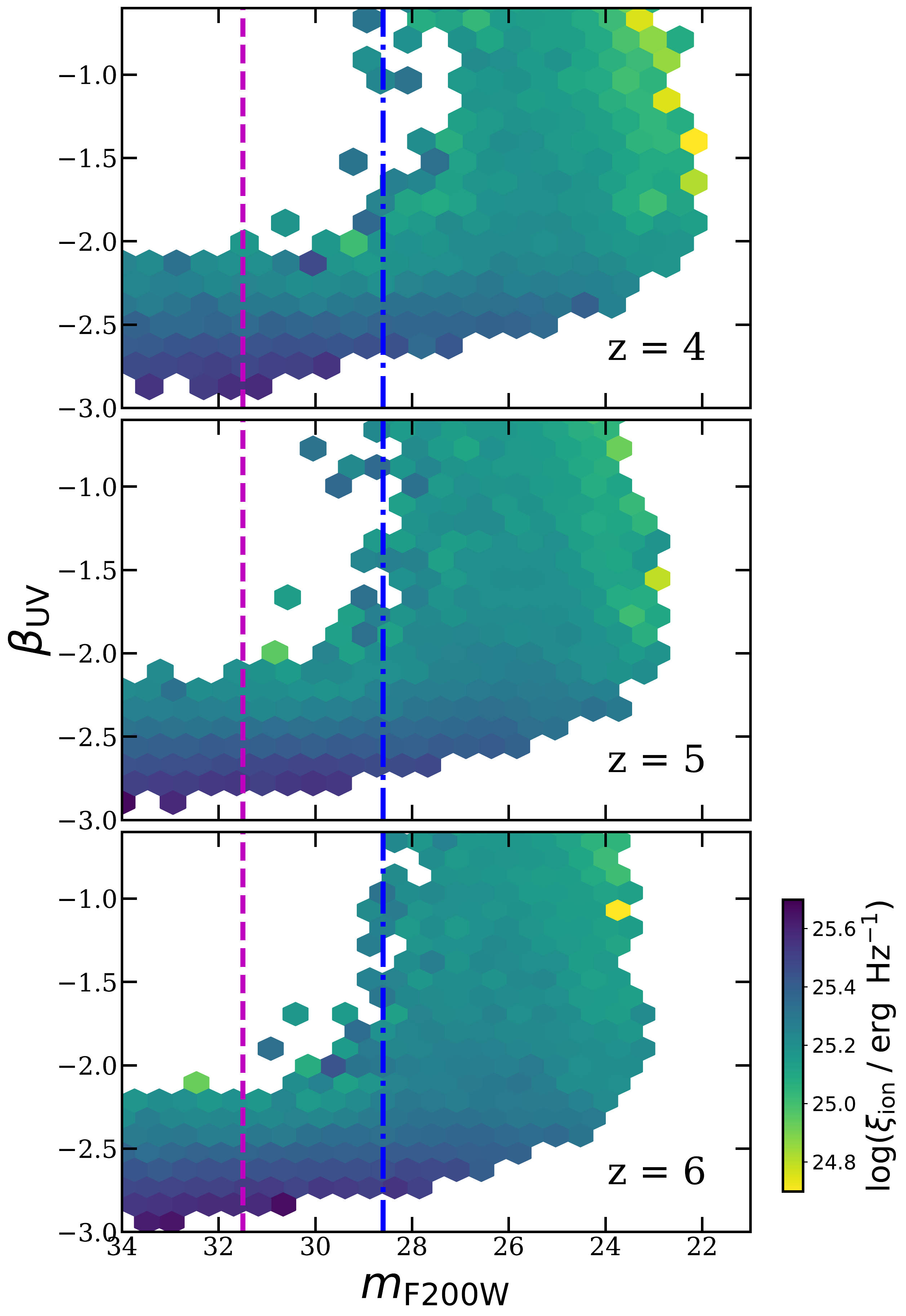}
    \caption{Correlation of $\beta_\text{UV}$ and $m_\text{F200W}$ at $z = 4$, 5, and 6 in our fiducial models, with the average value of \xion\ in each hexbin colour-coded. The vertical lines mark the detection limit of a typical \textit{JWST} NIRCam wide and deep surveys at $m_\text{F200W, lim} = 28.6$ and 31.5, respectively. The left edges of these plots coincide with the detection limit for lensed surveys.}
    \label{fig:3d_beta_F200W_stack} 
\end{figure}

\begin{figure}
    \includegraphics[width=\columnwidth]{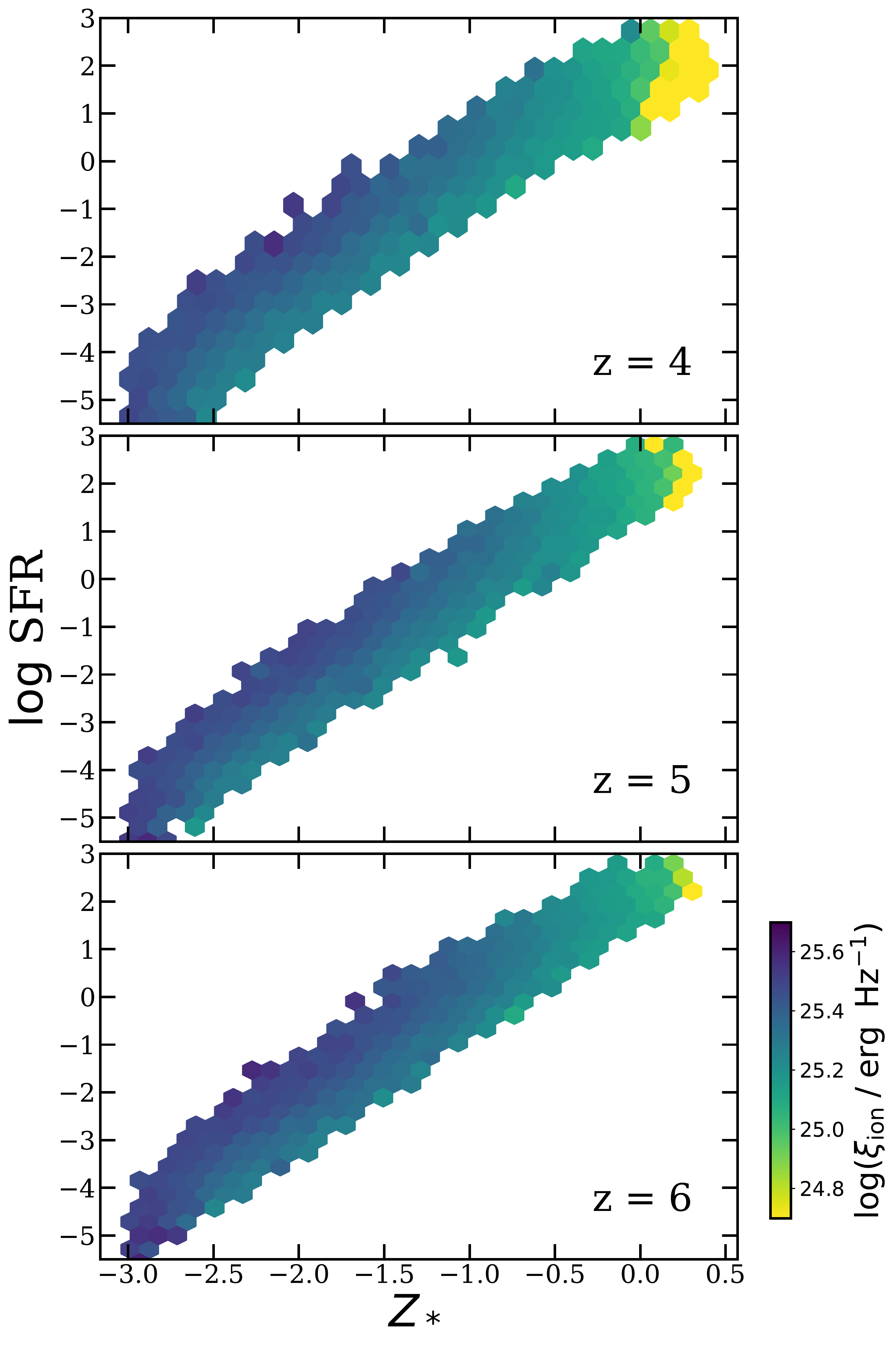}
    \caption{Correlation of stellar metallicity and SFR at $z = 4$, 5, and 6 in our fiducial models, with the average value of \xion\ in each hexbin colour-coded.}
    \label{fig:3d_SFR_Zstar_stack} 
\end{figure}

\citetalias{Finkelstein2019} obtained indirect constraints on the plausible range of \xion\ using an empirical model. They parametrization \xion\ as a function of redshift and rest-frame $M_\text{UV}$, and obtained the posterior on these parameters (along with other model parameters) using Markov Chain Monte Carlo (MCMC). They found that it was possible to satisfy constraints on the reionization history of the Universe with a relatively low average escape fraction (of about 5 percent) if \xion\ is higher in lower luminosity and higher redshift galaxies. The median posterior parameters found in their analysis are $d\log\xi_\text{ion}/dz = 0.13$ and $d\log\xi_\text{ion}/dz = 0.08$, and a reference magnitude of $M_\text{UV,ref} = 20$. Although our models predict these qualitative trends, our predicted trends are not nearly as strong as the ones suggested by the \citetalias{Finkelstein2019} models. We will investigate resolutions to this tension in more detail in \citetalias{Yung2020a} of this series.

Next we attempt to make predictions for correlations between physical and observable galaxy properties and \xion\ in order to suggest strategies for future observations. In fig. \ref{fig:3d_beta_F200W_stack}, we show observed-frame magnitude in the NIRCam F200W filter, $m_\text{F200W}$, vs. the UV-continuum slope, $\beta_\text{UV}$, in the $xy$-plane and colour code each cell by \xion. The detection limits for \textit{JWST} wide- and deep-field observations are also marked for quick reference, which are obtained assuming survey configurations that are comparable to those of past \textit{HST} surveys and the published NIRCam sensitivities with the F200W filter (see \citetalias{Yung2019}). We present these predictions for $z = 4$, 5, and 6, where many galaxies should be bright enough that spectra will be relatively efficiently obtained with \textit{JWST} or other facilities. We show a similar plot for sSFR and stellar metallicity in fig. \ref{fig:3d_SFR_Zstar_stack}.

\subsection{The impact of SN feedback}
\label{sec:alpha}

As shown in previous works in this series, the number density of faint, low-mass galaxies is very sensitive to the feedback strength, as stronger feedback suppresses star formation by ejecting gas from the ISM. However, the faint- and low-mass-end slope of UV LFs and SMF at high redshift are not well constrained by current observations. In \citetalias{Yung2019} and \citetalias{Yung2019a}, we experimented with alternative SN feedback slopes $\alpha_\text{rh}$, which effectively characterize the dependence of mass outflow rate on circular velocity, and found that a fiducial value of $\alpha_\text{rh} = 2.8$ yielded predictions that best match existing observational constraints. We further found that a range of values from $\alpha_\text{rh} = 2.0$ (weaker feedback) to $\alpha_\text{rh} = 3.6$ (stronger feedback) yielded predictions that are well within the current observational uncertainties. In this subsection, we explore how varying this model parameter impacts the production of ionizing photons.

\begin{figure}
    \includegraphics[width=\columnwidth]{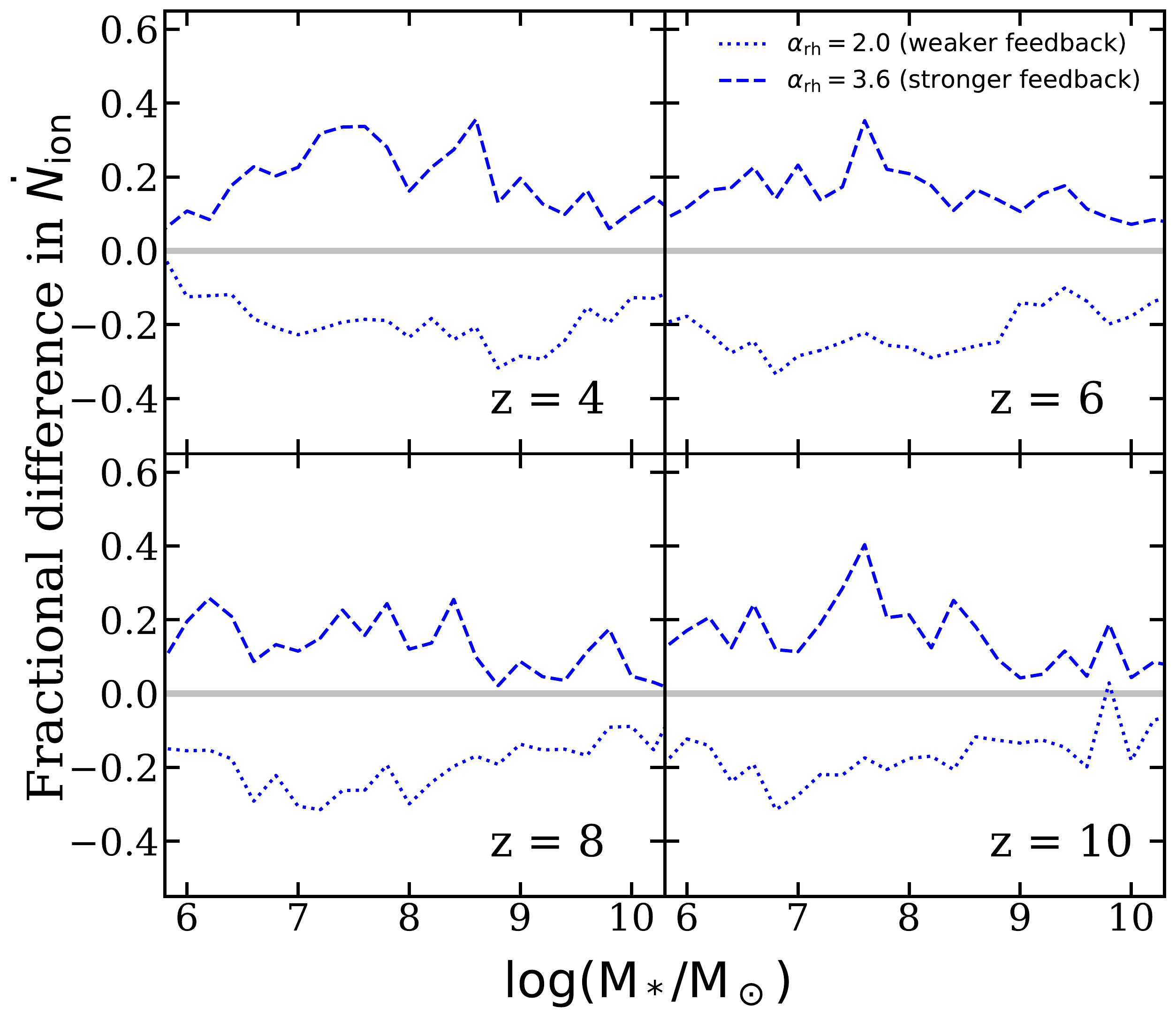}
    \caption{The fractional difference of the median of the distribution of \xion\ versus stellar mass for different SN feedback slopes predicted at $z=4$, 6, 8, and 10 with other model parameters and ingredients left at their fiducial values. The dashed and dotted blue lines represent the stronger and weaker feedback scenarios.}
    \label{fig:Nion_mstar_alp} 
\end{figure}

Fig. \ref{fig:Nion_mstar_alp} presents the fractional difference of the median of the predicted \Nion\ between our fiducial model ($\alpha_\text{rh} = 2.8$) and model variances with stronger ($\alpha_\text{rh} = 3.6$) and weaker ($\alpha_\text{rh}$ = 2.0) SN feedback, computed using a sliding boxcar filter of width $\Delta\log(M_*/\text{\Msun}) = 1.0$ in stellar mass. We find that the range of $\alpha_\text{rh}$ that yields galaxy populations within the observational uncertainties may cause $\sim20\%$ differences in \Nion, where more ionizing photons are produced when feedback is stronger and vice versa. This is likely caused by the more bursty star formation in the presence of stronger feedback.
Although stronger feedback may boost the production \emph{efficiency} of ionizing photons for low-mass galaxies, we must keep in mind that stronger feedback also suppresses the formation of low-mass galaxies, leading to lower overall UV luminosity density, so there is a trade-off.
We show in Fig. \ref{fig:productivity_z} and in our forthcoming \citetalias{Yung2020a} that in our models, stronger feedback leads to an overall decrease in the total number of ionizing photons. However, different implementations of stellar feedback could in principle produce different results.

\section{Discussion}
\label{sec:discussion}

In this work, we compute the production rate \Nion\ and production efficiency \xion\ of ionizing photons by combining stellar population synthesis models with a physically grounded model of galaxy formation set in a cosmological framework, the Santa Cruz SAM. Our fiducial model incorporates multiphase gas partitioning and \molh-based star formation recipes (\citetalias{Gnedin2011}-\citetalias{Bigiel2008}2), which have been shown to reproduce observational constraints over a wide range of redshifts from $z = 0$--10 (see \citetalias{Popping2014}, \citetalias{Somerville2015}, \citetalias{Yung2019}, and \citetalias{Yung2019a}). \Nion\ and \xion\ for composite stellar populations depends on the joint distribution of ages and metallicities in each galaxy. These quantities may in principle evolve with redshift or be correlated with global galaxy properties such as mass or luminosity. It is thus important to use a self-consistent, physically grounded model that hopefully contains the relevant internal correlations between physical properties. This is a significant advantage of our approach.

We compare \Nion\ and \xion\ calculated using different SPS models, which depends on the physical properties of the stellar populations, such as star formation and metal enrichment histories. This approach of modelling \Nion\ also makes physical predictions for the correlations among \xion\ and \Nion\ with other physical properties, such as $M_*$, $Z_*$, SFR, etc., and observable properties, such as $\beta_\text{UV}$ and observed frame magnitude. We also take advantage of the high efficiency of our model to explore how these predictions may be impacted by modelling uncertainties regarding SN feedback.

\subsection{Comparison with observations}

Our predictions for \xion\ are in good agreement with most of the observational measurements at $4 < z < 5$, but
despite the range of models we investigated in this work, none of our models reproduce the observed values of \xion\ at $z > 5$. Our models tend to predict values of \xion\ at $z\sim 7$ that are lower than the observational estimates by about 0.3 dex. Here we briefly discuss a few reasons that could give rise to this discrepancy. First, inferring \xion\ from observations is highly non-trivial. This estimate relies on obtaining the intrinsic rest-frame $M_\text{UV}$ using the UV slope $\beta_\text{UV}$ to correct for dust extinction. However, it is not known how reliable the empirical relations between UV slope and attenuation are at these redshifts, and uncertainties in the underlying shape of the attenuation curve also lead to uncertainties \citep{Bouwens2016a}.

In addition, due to limitations in current survey instruments, objects
with extremely bright emission lines are preferentially selected for
follow-up. These may be biased towards objects at the extreme tails of
the distribution of physical properties, or towards rare starbursting
galaxies. The large sample of lower redshift galaxies ($0 < z < 1$) studied by \citet{Izotov2017} show a broad range of values of \xion, and suggest the strong sensitivity in this parameter to recent bursts.

Finally, the estimates of \xion\ from the stellar population synthesis models are still quite uncertain. For example, the predicted values of \xion\ dropped by about 0.1 to 0.2 dex following changes to the stellar atmosphere models implemented in v2.1 of the \textsc{bpass} models. However, \citet{Eldridge2017} note that an improved treatment of rotation could increase \xion\ by a similar amount (see also \citet{Levesque2012}). Furthermore, \xion\ is quite sensitive to the high-mass end of the stellar IMF. For example, \citet{Eldridge2017} show that increasing the upper mass cutoff of the IMF from 100 to 300 \Msun\ increases the value of \xion\ by about 0.15 dex in the v2.1 \textsc{bpass} binary models.

\subsection{Our results in the context of other theoretical studies}

SAMs and hydrodynamic simulations are two different approaches to modelling galaxy formation (see \citet{Somerville2015a} for a detailed discussion), but they both require the use of phenomenological recipes for processes that cannot be explicitly simulated, such as star formation and stellar feedback. Both our models and the \textsc{BlueTides} simulations adopted similar multi-phase gas-partitioning and \molh-based star formation recipes, however, a number of other differences in the models may contribute to the discrepancies in their predictions. These include the choice of IMF, the quantities given priority in the calibration process, and the set of cosmological parameters adopted. \textsc{BlueTides} adopts cosmological parameters that are consistent with measured values reported by WMAP, which are different from the ones adopted in this work.
However, we do not expect this to lead to major differences in the relevant predicted galaxy properties (see discussion in Appendix C in \citetalias{Yung2019a}). In the model comparison presented in \citetalias{Yung2019} and \citetalias{Yung2019a}, we showed that the predicted statistical properties (rest-UV LFs, stellar mass functions, SFR functions) of galaxies at $z\sim 8$ -- 10 are nearly identical in our models and in \textsc{BlueTides}. However, when we compared our predicted values of \xion\ at a given stellar mass at $z\sim 8$ with those predicted by \citetalias{Wilkins2016a} for \textsc{BlueTides}, we find that the values agree well for the \citetalias{Bruzual2003} SPS models, while our predicted values are lower by about 0.1 dex for the single star \textsc{bpass} models and about 0.2 dex lower for the \textsc{bpass} models that include binary star evolution. We argue, based on the results presented in \citet{Eldridge2017}, that this difference is largely accounted for by changes in the \textsc{bpass} model results from the v2.0 version used by \citetalias{Wilkins2016a} to the v2.2.1 version that we have adopted in this work. This is supported by the fact that the predicted value of \xion\ from the FirstLight simulations (which also adopted the \textsc{bpass} v2.2.1 models) are quite similar to ours.

When we compare the other stellar population properties of galaxies in \textsc{BlueTides} and our models at the same stellar mass and redshift ($z=8$), we find that our galaxies are slightly younger and have higher specific star formation rates. By far the most significant difference, however, is that the stellar populations have mass-weighted metallicities that are a factor of six higher in our models than in \textsc{BlueTides}.

All of the physically grounded models in the literature (including \textsc{BlueTides}, FirstLight, and our models) predict rather mild redshift evolution in \xion\ as well as generally lower values than those implied by observations at $z\gtrsim 5$.

\subsection{Caveats and limitations of the modelling framework}

The productivity of ionizing photons of high-redshift galaxies connects the small-scale processes of stellar evolution, star formation, and stellar feedback with cosmological scale processes such as the reionization of intergalactic hydrogen.
Our results depend critically on stellar population synthesis models, which have their own limitations and uncertainties \citep{Conroy2013}. The massive, low-metallicity stars that are the most efficient at producing ionizing photons are particularly sensitive to modelling uncertainties in physical processes, such as mass loss, mixing, convection, rotation, and magnetic fields; and significant progress has been made in the past few years on improving these models. One area that has received some attention within the literature on high redshift galaxy populations and the epoch of reionization recently is the impact of binary processes on the evolution of massive stars and the implications for the production of ionizing photons \citep{Wilkins2016a,Ma2016,Ceverino2019}. Binarity can have a significant effect on the production rate of ionizing photons via a diverse range of physical mechanisms. Mass transfer and mergers among binaries can increase the number of massive stars present over times longer than the lifetime of massive single stars. In addition, mass transfer can produce rapidly rotating `chemically homogeneous' stars, which are more prevalent at low metallicity and are highly efficient at producing ionizing photons. The rotating massive stellar population can be an significant source of ionizing photons in the first $\sim4$ Myr for starburst galaxies \citep{Choi2017a}. A third process is that stars in a binary system can be stripped of their envelope, exposing the hot core of the star. Stripped stars can boost the production rate of ionizing photons at times 20--500 Myr after the onset of star formation by as much as 1--2 orders of magnitude \citet{Gotberg2019}. The \textsc{bpass} models adopted in this work include modelling of these processes \citep{Eldridge2011,Eldridge2012,Stanway2016,Eldridge2017}, while the role of stripped stars has been highlighted in \citet{Gotberg2019}. Other processes that may be important for producing ionizing photons, but which are not accounted for or not modelled in as much detail in the \textsc{bpass} models, include accreting white dwarfs and X-ray binaries \citep{Madau2017a}. Because \xion\ is so sensitive to the most massive stars, it is also quite sensitive to details of the high-mass slope and cutoff of the stellar IMF \citep[e.g.][]{Eldridge2017}. In summary, the raw predictions of \xion\ even in the latest models seem to be uncertain at the level of at least a factor of two.

Moving up in scale, the star formation and chemical enrichment histories that form the backbone of this work are based on the modelling framework developed in a series of papers including \citetalias{Somerville2015}, \citetalias{Yung2019}, and \citetalias{Yung2019a}. These models make a large number of simplifying assumptions about how baryonic processes in galaxies operate. A fundamental ansatz of our models is that the scaling relations that describe processes such as gas partitioning, star formation efficiency, and stellar feedback are valid over all of cosmic time. This may not necessarily be the case. Our adopted scaling relations may have dependencies that do not properly track the true physical properties, which may evolve differently. These assumptions are typically tested and calibrated by comparing with observations. As shown in these previous works, the predicted distribution functions for $M_\text{UV}$, $M_*$, and SFR up to $z\sim10$ are in very good agreement with observations in regimes where observational constraints are available. However, the physical properties of high-redshift faint galaxies, as well as the underlying physical processes that drive them, are poorly constrained.

The results of this work hint that the treatment of stellar feedback and chemical evolution may be an area that requires careful examination in future work.
Although our models produce excellent agreement with both the \textsc{BlueTides} and FIRE simulations with regard to bulk properties such as the stellar-to-halo-mass ratio, as noted above, the predicted mass-metallicity relation at $z\sim 6$--8 is substantially different.
As shown in Fig. 7 of \citet{Ma2016a} and Fig. 6 of \citet{Somerville2015a}, the predicted mass-metallicity relation shows a large dispersion between different models and simulations. This is because there are many details of the chemical enrichment process that are poorly understood or simplified in cosmological galaxy formation models. For example, the version of the SC-SAM models used here assumes instantaneous recycling, i.e., that metals ejected by SNe are instantaneously mixed with the whole ISM. Our models assume a fixed chemical yield, $y$, which is treated as a fixed parameter that is tuned such that the predicted stellar mass-metallicity relation matches $z\sim0$ observations. We also assume that the metallicity of outflows is the same as that of the ISM, while there is evidence in nearby galaxies that stellar driven winds may be substantially metal-enhanced  \citep[metals preferentially ejected][]{Chisholm2018}.
Furthermore, we note that the multiphase-gas partitioning recipes adopted in our models may break down in extremely metal-poor environments \citep{Sternberg2014}. Neither primordial \molh\ cooling and Pop III stars nor metal enrichment by these objects is explicitly included in our models. Instead, top-level progenitor halos are polluted to a metallicity floor with typical values of $Z_\text{pre-enrich} = 10^{-3} Z_\odot$.

\subsection{Characterizing ionizing sources with JWST and beyond}
As shown in fig. \ref{fig:3d_beta_F200W_stack}, predictions from our model show that $\beta_\text{UV}$ may not be an very good tracer for \xion\ as the correlation between quantities is rather weak.
The productivity of ionizing photons can be indirectly traced with a number of spectral features. Due to current instrumental limitations, it is tremendously challenging to obtain high-resolution spectra for galaxies at high redshifts. The NIRSpec onboard \textit{JWST} possesses the advanced capability for detecting H$\alpha$ at $z\lesssim7$, and given adequate exposure time, the Mid-Infrared Instrument (MIRI) will also be able to pick up signals from bright objects at $z\gtrsim7$. Furthermore, planned flagship ground-based telescopes, such as the Extremely Large Telescope (ELT), the Thirty Meter Telescope (TMT), and the Giant Magellan Telescope (GMT), are also expected to enable these kinds of observations for many more objects and for fainter, perhaps more `typical' galaxies. These anticipated observations will be able to constrain or even rule out completely some of the predictions provided in this work.

\subsection{Implications of our results for cosmic reionization}

One of the main goals of this work is to provide a physical explanation for how the intrinsic production efficiency of ionizing photons scales with physical properties of galaxies and how these relations have evolved during the EoR. A very common assumption in reionization modelling \citep[e.g.][]{Robertson2013} is that the cosmic ionizing emissivity can be written as
\begin{equation}
    \dot{n}_{\rm ion} = f_{\rm esc} \xi_{\rm ion} \rho_{\rm UV}
\end{equation}
i.e., implicitly assuming that $f_{\rm esc}$ and $\xi_{\rm ion}$ have a fixed value with no scatter, that does not change over cosmic time. While this may be true in an effective, integrated sense, in this series of papers, we have developed more realistic models in which both of these parameters may be correlated with internal galaxy properties, such as their UV luminosity, stellar mass, or metallicity. These properties, in turn, are expected to be correlated with galaxy clustering and large scale environment, which likely has implications for reionization. We are building towards a model framework that can efficiently go from dark matter properties and a set of assumptions about the physical processes involved in galaxy formation, to a cosmic reionization history and ultimately to predicted maps of observable tracers of the reionization.

\begin{figure}
    \includegraphics[width=\columnwidth]{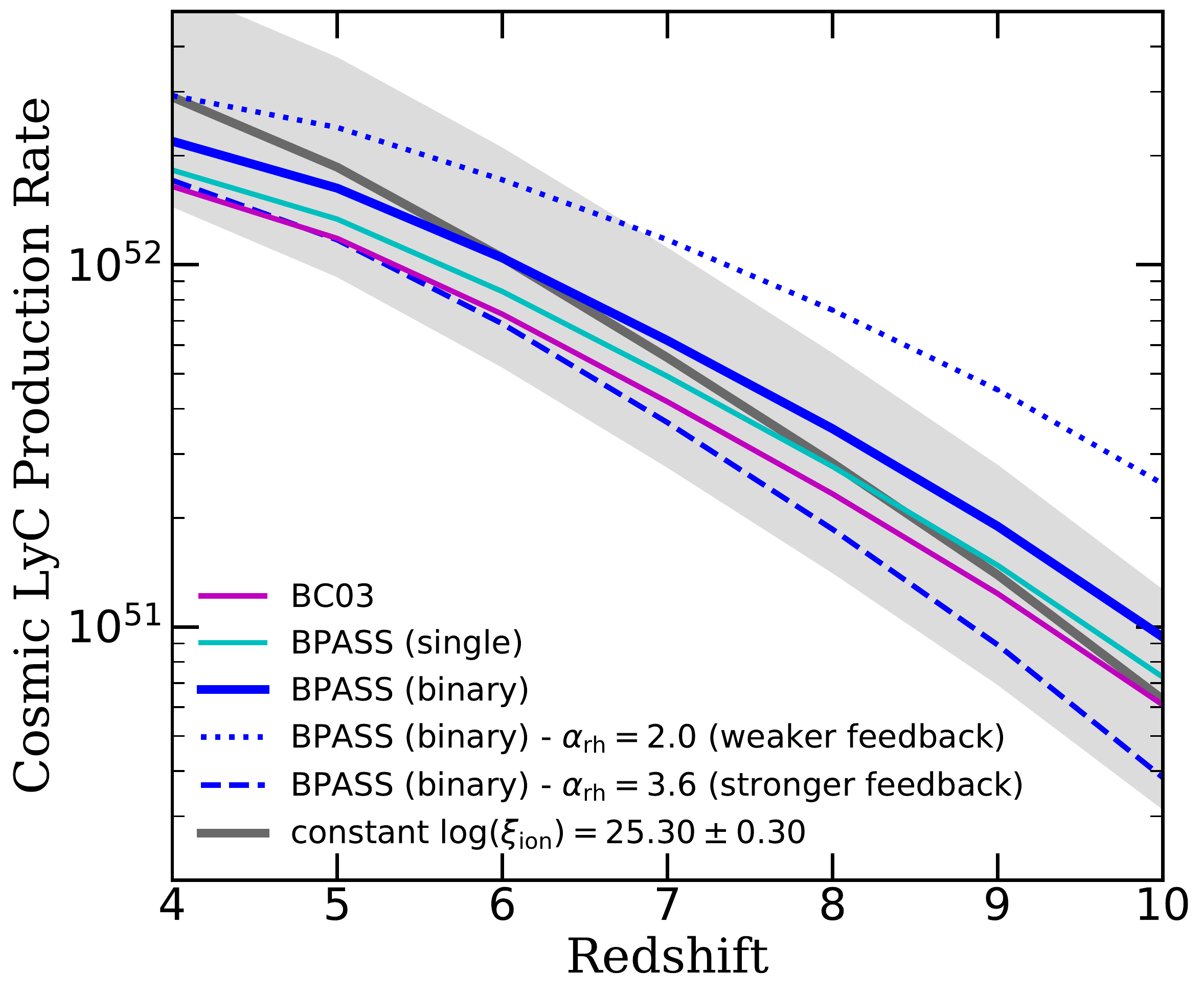}
    \caption{Cosmic LyC photon production rate evolution with redshift given in units of Mpc$^{-3}$ s$^{-1}$, calculated using SPS models from \citetalias{Bruzual2003} (purple) and \textsc{bpass} (cyan and blue for single and binary). For different SN feedback slopes predicted at $z = 4$ -- 10 with other model parameters and ingredients left at their fiducial values, the dashed and dotted blue lines represent the stronger and weaker feedback scenarios, respectively. The grey lines show the case where $\log(\xi_\text{ion}) = 25.30 \pm 0.30$.}
    \label{fig:productivity_z} 
\end{figure}

A summary of our predictions for the \emph{intrinsic} cosmic ionizing emissivity is shown in Fig. \ref{fig:productivity_z}. This figure illustrates the impact of adopting different stellar population models (self-consistently incorporated within our models) and of varying the strength of stellar feedback. As discussed above, the rather strong metallicity dependence of \xion\ implies that the details of the interplay between chemical evolution and stellar populations can lead to differences of up to factors of a few, which could be significant for reionization. The total number of ionizing photons produced are degenerately affected by the abundance of galaxies during the EoR and the intrinsic productivity of ionizing radiation. As shown in \citetalias{Yung2019} and \citetalias{Yung2019a}, the abundances of low-mass galaxies are very sensitive to the strength of SN feedback and the number of objects predicted can span about an order of magnitude for different model assumptions. On the other hand, in \S3.4 we also show that stronger feedback yields slightly ($\sim 20\%$) boosted ionizing photon production efficiencies.

Whether galaxies alone are sufficient to ionize the IGM remains a fundamental open question. Historically, there has been tension between the total ionizing photon budget accounted for by galaxies, and constraints on the onset and duration of the EoR. Analytic calculations \citep[e.g.][]{Kuhlen2012, Robertson2015} have shown that galaxies are likely to have produced sufficient numbers of ionizing photons to reionize the Universe under the condition that faint objects remain fairly numerous below the current detection limit and are fairly efficient at producing ionizing photons. In \citetalias{Yung2019} and \citetalias{Yung2019a}, we showed that in our models, the faint-end slope of the rest-frame UV LF remains steep down to $M_\text{UV}\sim -9$, under the assumption that SNe feedback at high redshifts operates with similar efficiency as that required to match $z\sim0$ observations. The predicted ionizing photon production efficiencies, especially with models that include binary evolution, are well within the range $\log(\xi_\text{ion}) \sim 25.30 \pm 0.30$ (or higher) adopted in recent models that are able to plausibly account for observable constraints on the reionization history \citep{Robertson2013,Robertson2015,Kuhlen2012,Finkelstein2019}. Therefore, we can already anticipate that our models will also produce good agreement with the observed reionization history, subject to uncertainties on the escape fraction of ionizing radiation. We present a detailed analysis of the implications for the cosmic reionization history of our full modelling framework, including the escape fraction, in \citetalias{Yung2020a}.

\section{Summary and conclusions}
\label{sec:snc}

In this work, we implemented a physically motivated approach to calculating \Nion\ in the Santa Cruz SAM, accounting for stellar age and metallicity distribution of the stellar population in each galaxy. We present results predicted based on SSP models from \citetalias{Bruzual2003} and both single and binary stellar population by \textsc{bpass}. Taking advantage of the high efficiency of our modelling framework, we make predictions for galaxies across $z=4$--10, forming in a wide range of halo masses, ranging from $V_\text{vir}\approx20$--500\kms. These predictions provide forecasts for future observations with \textit{JWST} and also put modelling constraints on the role high-redshift, low-mass galaxies played throughout the EoR. We also provide predicted scaling relations of \Nion\ with $M_*$, $M_\text{h}$, $\beta_\text{UV}$, and other galaxy properties. We compared our predictions to other models from the literature and the latest observations in both the intermediate- and high-redshift universe.

We summarize our main conclusions below.
\begin{enumerate}
    \item In agreement with previous work, we find that SPS models accounting for binary stellar populations can produce a factor of $\sim2$ more ionizing radiations than models that only account for single star populations for galaxies across all mass ranges.
    
    \item We find that faint, low-mass galaxies can have ionizing efficiencies that are up to a factor of two higher than those of bright, massive ones. This is due to the strong correlation between stellar mass (or luminosity) and metallicity that is already in place at early times in our models, and the strong dependence of \xion\ on metallicity, particularly in models that include physical processes related to stellar binaries.
    
    \item Our models predict a rather weak dependence of \xion\ with redshift (about $\sim 0.23$ dex or factor of $\sim 0.6$ decrease from $z\sim12$ to 4).
    
    \item We find that increasing of the strength of SN feedback may allow galaxies to produce $\sim 20\%$ more ionizing photons, because their star formation is more bursty. However, we also note that stronger feedback suppresses star formation in low-mass halos and decreases the number density of low-mass galaxies, which ultimately decreases the total number of ionizing photons.
    
    \item Our predicted median values of \xion\ are significantly lower than all current observational estimates at $z > 5$. This may be due to sampling bias towards extremely bright and bursty objects in current high-redshift observations, or to modelling uncertainties in stellar population synthesis models.
    
    \item We provide scaling relations for \Nion\ as a function of $M_\text{h}$ and a number of other galaxy properties, which maybe useful for cosmological scale reionization simulations where baryonic physics is not resolved.
    
\end{enumerate}

\section*{Acknowledgements}
The authors of this paper would like to thank Adriano Fontana et al. for organizing the `The Growth of Galaxies in the Early Universe -- V' conference in Sesto, Italy, which fostered significant discussions that inspired the creation of this work. We also thank Selma de Mink and Stephen Wilkins for useful comments and discussions. We also thank the anonymous referee for the constructive comments that improved this work. AY and RSS thank the Downsbrough family for their generous support, and gratefully acknowledge funding from the Simons Foundation.


\bibliographystyle{mnras}
\bibliography{library.bib} 


\appendix

\section{Tabulated values for ionizing photon productivity}
\label{appendix:a}
\setcounter{table}{0} \renewcommand{\thetable}{A\arabic{table}}

Tabulated $\log(\dot{N}_\text{ion}/M_\text{h})$ from our fiducial model is provided in Table \ref{table:tab_NionMH}. Other scaling relations shown in this work are available online.

\begin{table*}
    \centering
    \caption{Tabulated $\log(\dot{N}_\text{ion}/M_\text{h})$ at $z = 4$--10 from our fiducial model. These are the median values as shown in fig. \ref{fig:NionMH_bin}. The upper and lower limits give represent the value of the 84th and the 16th percentile.}
    \label{table:tab_NionMH}
    \scalebox{0.95}{
        \begin{tabular}{cccccccccc}
        \hline
        & \multicolumn{7}{c}{$\log_{10}((\dot{N}_\text{ion}/M_\text{h})\; /\; (s\;M_\odot)^{-1})$} \\
        $\log_{10}(M_\text{h}/\text{\Msun})$ & $z = 4$ & $z = 5$  & $z = 6$  & $z = 7$  & $z = 8$  & $z = 9$  & $z = 10$   \\
        \hline
        8.25 & 39.17$^{+0.47}_{-0.70}$ & 39.62$^{+0.54}_{-0.66}$ & 40.04$^{+0.57}_{-0.65}$ & 40.46$^{+0.52}_{-0.66}$ & 41.07$^{+0.40}_{-0.72}$ & 41.29$^{+0.31}_{-0.60}$ & 41.45$^{+0.24}_{-0.54}$ \\ [0.1in]
        8.50 & 39.68$^{+0.42}_{-0.66}$ & 40.48$^{+0.42}_{-0.79}$ & 40.93$^{+0.31}_{-0.66}$ & 41.21$^{+0.26}_{-0.47}$ & 41.34$^{+0.31}_{-0.45}$ & 41.55$^{+0.25}_{-0.55}$ & 41.68$^{+0.23}_{-0.53}$ \\ [0.1in]
        8.75 & 40.46$^{+0.32}_{-0.72}$ & 40.94$^{+0.28}_{-0.60}$ & 41.22$^{+0.30}_{-0.50}$ & 41.41$^{+0.31}_{-0.45}$ & 41.64$^{+0.27}_{-0.48}$ & 41.77$^{+0.27}_{-0.54}$ & 41.89$^{+0.25}_{-0.60}$ \\ [0.1in]
        9.00 & 40.88$^{+0.30}_{-0.66}$ & 41.22$^{+0.32}_{-0.59}$ & 41.45$^{+0.31}_{-0.47}$ & 41.68$^{+0.28}_{-0.49}$ & 41.85$^{+0.27}_{-0.46}$ & 42.02$^{+0.25}_{-0.57}$ & 42.12$^{+0.26}_{-0.60}$ \\ [0.1in]
        9.25 & 41.21$^{+0.33}_{-0.61}$ & 41.48$^{+0.31}_{-0.50}$ & 41.72$^{+0.33}_{-0.45}$ & 41.92$^{+0.28}_{-0.43}$ & 42.10$^{+0.27}_{-0.55}$ & 42.27$^{+0.23}_{-0.49}$ & 42.38$^{+0.21}_{-0.66}$ \\ [0.1in]
        9.50 & 41.45$^{+0.33}_{-0.54}$ & 41.72$^{+0.33}_{-0.46}$ & 41.97$^{+0.28}_{-0.44}$ & 42.16$^{+0.28}_{-0.51}$ & 42.31$^{+0.26}_{-0.55}$ & 42.47$^{+0.23}_{-0.56}$ & 42.61$^{+0.20}_{-0.55}$ \\ [0.1in]
        9.75 & 41.68$^{+0.39}_{-0.47}$ & 41.97$^{+0.34}_{-0.46}$ & 42.20$^{+0.28}_{-0.45}$ & 42.36$^{+0.26}_{-0.40}$ & 42.57$^{+0.20}_{-0.44}$ & 42.72$^{+0.15}_{-0.44}$ & 42.83$^{+0.16}_{-0.52}$ \\ [0.1in]
        10.00 & 41.89$^{+0.36}_{-0.47}$ & 42.18$^{+0.30}_{-0.39}$ & 42.42$^{+0.25}_{-0.38}$ & 42.60$^{+0.22}_{-0.42}$ & 42.77$^{+0.18}_{-0.35}$ & 42.88$^{+0.17}_{-0.39}$ & 43.01$^{+0.13}_{-0.50}$ \\ [0.1in]
        10.25 & 42.09$^{+0.37}_{-0.45}$ & 42.38$^{+0.29}_{-0.38}$ & 42.60$^{+0.25}_{-0.32}$ & 42.78$^{+0.21}_{-0.36}$ & 42.95$^{+0.16}_{-0.40}$ & 43.05$^{+0.16}_{-0.42}$ & 43.16$^{+0.17}_{-0.43}$ \\ [0.1in]
        10.50 & 42.32$^{+0.31}_{-0.40}$ & 42.60$^{+0.24}_{-0.36}$ & 42.82$^{+0.21}_{-0.35}$ & 42.98$^{+0.18}_{-0.31}$ & 43.14$^{+0.16}_{-0.38}$ & 43.22$^{+0.17}_{-0.37}$ & 43.28$^{+0.20}_{-0.43}$ \\ [0.1in]
        10.75 & 42.54$^{+0.29}_{-0.44}$ & 42.79$^{+0.24}_{-0.37}$ & 43.06$^{+0.17}_{-0.42}$ & 43.15$^{+0.17}_{-0.39}$ & 43.28$^{+0.17}_{-0.43}$ & 43.36$^{+0.17}_{-0.45}$ & 43.42$^{+0.20}_{-0.47}$ \\ [0.1in]
        11.00 & 42.81$^{+0.22}_{-0.49}$ & 43.06$^{+0.15}_{-0.47}$ & 43.23$^{+0.13}_{-0.39}$ & 43.34$^{+0.13}_{-0.41}$ & 43.44$^{+0.16}_{-0.43}$ & 43.47$^{+0.19}_{-0.41}$ & 43.55$^{+0.20}_{-0.54}$ \\ [0.1in]
        11.25 & 43.04$^{+0.14}_{-0.57}$ & 43.22$^{+0.13}_{-0.43}$ & 43.35$^{+0.13}_{-0.46}$ & 43.47$^{+0.14}_{-0.41}$ & 43.47$^{+0.21}_{-0.40}$ & 43.55$^{+0.22}_{-0.48}$ & 43.61$^{+0.21}_{-0.50}$ \\ [0.1in]
        11.50 & 43.18$^{+0.10}_{-0.50}$ & 43.34$^{+0.14}_{-0.44}$ & 43.43$^{+0.15}_{-0.39}$ & 43.51$^{+0.17}_{-0.38}$ & 43.59$^{+0.19}_{-0.45}$ & 43.61$^{+0.22}_{-0.46}$ & 43.60$^{+0.28}_{-0.54}$ \\ [0.1in]
        11.75 & 43.23$^{+0.15}_{-0.45}$ & 43.37$^{+0.15}_{-0.47}$ & 43.44$^{+0.17}_{-0.33}$ & 43.52$^{+0.17}_{-0.33}$ & 43.54$^{+0.20}_{-0.41}$ & 43.54$^{+0.23}_{-0.42}$ & 43.53$^{+0.25}_{-0.47}$ \\ [0.1in]
        12.00 & 43.13$^{+0.26}_{-0.60}$ & 43.24$^{+0.21}_{-0.36}$ & 43.31$^{+0.20}_{-0.30}$ & 43.38$^{+0.19}_{-0.37}$ & 43.42$^{+0.20}_{-0.34}$ & 43.46$^{+0.21}_{-0.36}$ & 43.40$^{+0.29}_{-0.44}$ \\ [0.1in]
        12.25 & 42.80$^{+0.28}_{-1.11}$ & 43.00$^{+0.18}_{-0.28}$ & 43.12$^{+0.21}_{-0.31}$ & 43.22$^{+0.23}_{-0.32}$ & 43.33$^{+0.18}_{-0.34}$ & 43.38$^{+0.21}_{-0.43}$ & 43.34$^{+0.24}_{-0.33}$ \\ [0.1in]
        \hline
        \end{tabular}
        }
\end{table*}

\bsp	
\label{lastpage}
\end{document}